\documentclass[superscriptaddress,pra,aps,twocolumn,10pt,longbibliography]{revtex4-2}
\usepackage{bm}
\usepackage{amsmath}
\usepackage{amsfonts}
\usepackage{amssymb}
\usepackage{physics}
\usepackage{graphicx}
\usepackage{grffile}
\usepackage{natbib}
\usepackage{epstopdf}
\usepackage[pdftex]{color}
\usepackage[dvipsnames]{xcolor}
\usepackage{hyperref}
\usepackage{verbatim}
\usepackage{float}
\usepackage[boxed]{algorithm}
\usepackage{algpseudocode}
\hypersetup{
    colorlinks=true,       
    linkcolor=cyan,          
    citecolor=magenta,        
    filecolor=magenta,      
    urlcolor=cyan,           
    runcolor=cyan
}
\usepackage[normalem]{ulem}

\newcommand{\beq}{\begin{eqnarray}}
\newcommand{\eeq}{\end{eqnarray}}
\newcommand{\bes} {\begin{subequations}}
\newcommand{\ees} {\end{subequations}}

\newcommand{\ignore}[1]{}

\graphicspath{{figs/}}

\begin{document}

\title{Comparing relaxation mechanisms in quantum and classical transverse-field annealing}

\author{Tameem Albash}
\affiliation{Department of Electrical and Computer Engineering,  University of New Mexico, Albuquerque, New Mexico 87131, USA}
\affiliation{Department of Physics and Astronomy and Center for Quantum Information and Control, CQuIC, University of New Mexico, Albuquerque, New Mexico 87131, USA}

\author{Jeffrey Marshall}
\thanks{Authors contributed equally}
\email{talbash@unm.edu}
\email{jmarshall@usra.edu}
\affiliation{QuAIL, NASA Ames Research Center, Moffett Field, California 94035, USA}
\affiliation{USRA Research Institute for Advanced Computer Science, Mountain View, California 94043, USA}

\begin{abstract}
Annealing schedule control provides new opportunities to better understand the manner and mechanisms by which putative quantum annealers operate.
By appropriately modifying the annealing schedule to include a pause (keeping the Hamiltonian fixed) for a period of time, we show it is possible to more directly probe the dissipative dynamics of the system at intermediate points along the anneal and examine thermal relaxation rates, for example, by observing the re-population of the ground state after the minimum spectral gap. We provide a detailed comparison of experiments from a D-Wave device, simulations of the quantum adiabatic master equation and a classical analogue of quantum annealing, spin-vector Monte Carlo, and we observe qualitative agreement, showing that the characteristic increase in success probability when pausing is not a uniquely quantum phenomena.  We find that the relaxation in our system is dominated by a single time-scale, which allows us to give a simple condition for when we can expect pausing to improve the time-to-solution, the relevant metric for classical optimization.
Finally, we also explore in simulation the role of temperature whilst pausing as a means to better distinguish quantum and classical models of quantum annealers.
\end{abstract}
\maketitle

\section{Introduction}
Over the last few years there has been a healthy debate surrounding the operation of current generation quantum annealers, such as the D-Wave family of devices \cite{Dwave,Harris:2010kx,Bunyk:2014hb}.  These devices implement in hardware a realization of quantum annealing \cite{finnila_quantum_1994,Brooke1999,kadowaki:98,farhi:01,santoro:02}, whereby a continuous-time interpolation between two non-commuting Hamiltonians is performed.  At any point along the interpolation, the low-energy spectrum of the device is approximated by a transverse field Ising model, with the relative strength of the transverse field and Ising Hamiltonians determined by the interpolation schedule. 
%

These devices are known to be sensitive to the operating temperature \cite{marshall-rieffel-hen-2017}, although the exact model for this is still under scrutiny. 
On the one hand, some experimental observations (e.g.~\cite{king-topological,Harris162}) are consistent with a quasi-static model \cite{amin-freezeout} induced by the coupling of the quantum system to a non-zero temperature bath \cite{adiabaticME}, but on the other, studies aiming to directly observe quantum thermal distributions have been mixed \cite{powerOfPausing, izquierdo-thermal-sampler,lanting:14,Harris162,king-topological,2020arXiv200301019K,2020arXiv200710555K}. Whilst these previous results demonstrate clearly that the dynamics associated with pausing are purely dissipative, whether these processes are able to improve performance in the context of optimization, where the goal is to reach or approximate the ground state of the Ising Hamiltonian, and ultimately help achieve an advantage over classical approaches is still unknown. 

The attempt to measure non-trivial quantum thermal statistics on such devices utilizes a mid-anneal pause, whereby the Hamiltonian is held fixed such that the system can  equilibrate. Given a sufficiently fast quench (\emph{or} ramp) and read-out, it should in principle be possible to observe the quantum Gibbs distribution associated with the Hamiltonian at the pause. To date, numerical studies into the pause have been conducted purely by quantum adiabatic master equation (AME) simulations \cite{passarelli-pause, passarelli-reverse, chen-pause}, which generally agree with observations in experiments \cite{powerOfPausing}. In Ref.~\cite{chen-pause}, a deeper study into the theory of pausing was conducted, from the point of view of quantum thermalization. Here sufficient conditions were identified such that a pause has a non-trivial influence on the output of an anneal. These results are consistent with the picture put forward in Ref.~\cite{powerOfPausing} whereby the competition of various timescales -- relaxation, pause, annealing -- determines the general dynamics during the mid to late anneal region.  

These timescales are not unique to the quantum AME however, motivating a study to determine whether or not the ground state statistics under pausing can be reproduced by a simple semi-classical model of the quantum annealing dynamics.
In such a picture, the quantum Hamiltonian is replaced by a related classical potential \cite{PhysRevD.19.2349,PhysRevA.94.062106}, which is then explored by thermal hopping (with no quantum tunneling). Here we use spin-vector Monte Carlo (SVMC) \cite{svmc} as our model, where qubits are replaced by 2-dimensional rotors (with no entanglement), and the system evolves using a Metropolis-Hastings algorithm \cite{10.1063/1.1699114,10.1093/biomet/57.1.97}, whereby each rotor's orientation is updated at some fixed temperature while the potential is changed as an analog to the Hamiltonian changing during the quantum anneal. Such a description has been extremely successful in discriminating quantum from classical effects in such devices and to what extent entanglement determines the output statistics \cite{svmc,Albash:14,Albash:2014if,albash-2015-entanglement}.

By adapting the standard update in SVMC, we are able to capture the general quasi-static (thermal) behavior observed in experiments on D-Wave devices, using the pause feature as a probe for the dynamics.
This adaptation to the original SVMC algorithm, we believe, will be helpful in understanding thermal dynamics more generally in such devices going forward. Such a picture provides a simple intuition for the role of thermal effects during the anneal and whilst pausing, and can help determine whether pausing can be expected to provide an advantage over classical thermal methods for sampling.

Motivated by our experimental and simulation results, we derive a condition under which the time-to-solution (TTS) can be reduced by relaxation effects. This condition is agnostic to the underlying dynamics and only depends on the observation of a single dominant time scale determining the exponential re-population effect associated with the pause. Analysis of this type can be useful in determining whether pausing can improve performance.

Our paper is organized as follows. In Sec.~\ref{sec:Background}, we provide details about the problem instance that we focus on in this study.  In Sec.~\ref{sec:Methods}, we describe our simulation methods.  In Sec.~\ref{sec:Results}, we give our experimental results using the D-Wave 2000Q device and our simulation results, and in Sec.~\ref{sec:Condition} we provide a simple condition under which we can expect pausing to improve the performance of quantum annealing in the context of classical optimization.  We conclude with a discussion of our results in Sec.~\ref{sec:Conclusions}.

\section{Background} \label{sec:Background}
We study the standard transverse field Ising model of the form $H(s) = A(s)H_x + B(s)H_p$ where the transverse field Hamiltonian $H_x=-\sum_i \sigma_i^x$ is known as the driver Hamiltonian, and the Ising Hamiltonian $H_p = \sum_{i<j}J_{ij}\sigma_i^z \sigma_j^z + \sum_i h_i \sigma_i^z$ is the classical `problem' Hamiltonian. The functions $A(s), B(s)$ determine the interpolation between $H_x$ and $H_p$, and the dimensionless annealing parameter $s \in [0,1]$ is a function of the physical time $t$.  For the standard anneal, it is given by $s(t) = t/t_a$, where $t_a$ denotes the total annealing time, but in general it can be any piece-wise continuous function of $t$. 

\begin{figure}
    \centering
    \includegraphics[width=0.98\columnwidth]{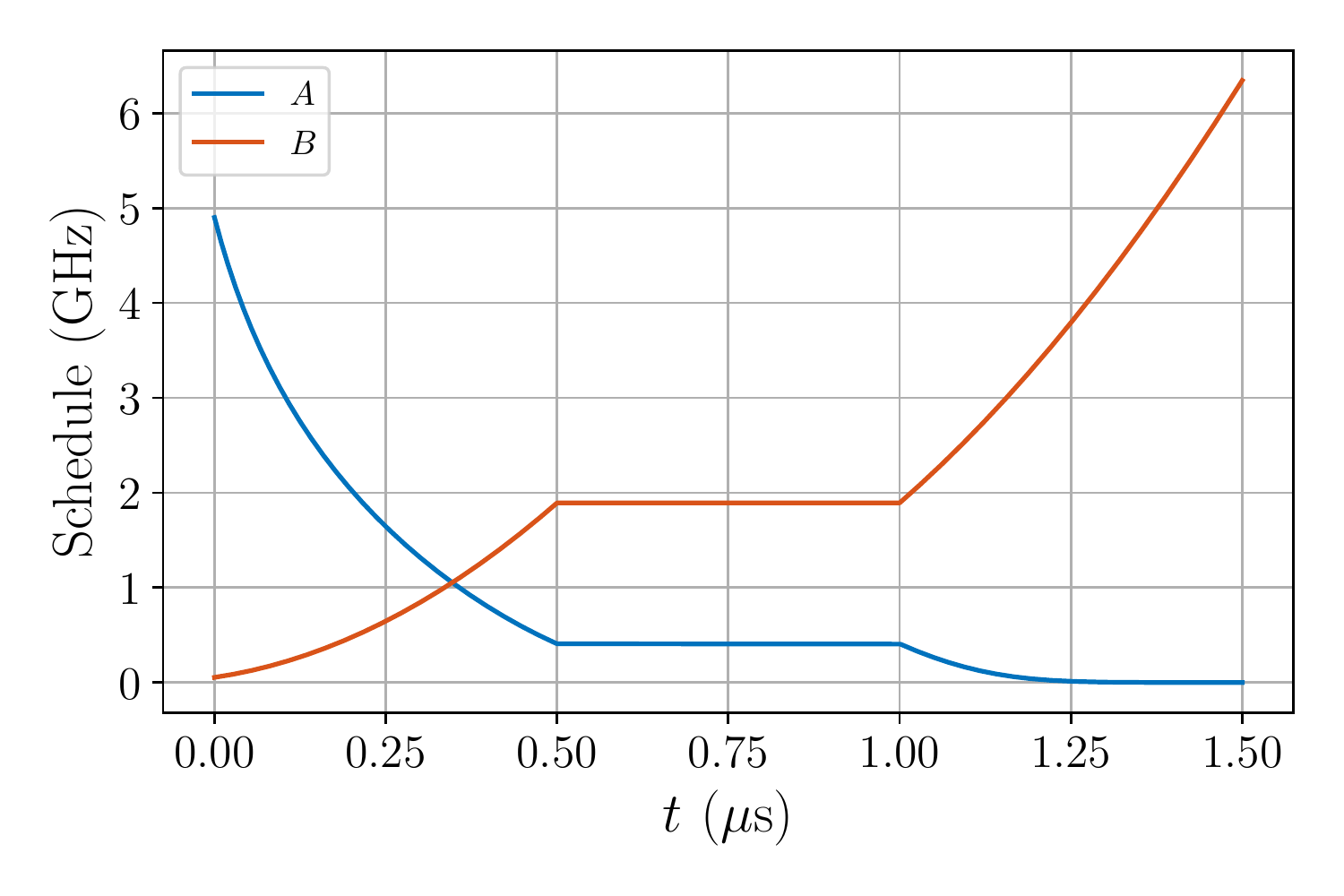}
    \caption{Example schedule with a pause of length $t_p=0.5\mu$s  inserted into an anneal with anneal time $t_a=1\mu$s. The pause occurs at $s_p=0.5$ (i.e. midway through the interpolation).  We sometimes refer to the anneal after the pause as the ramp. Units of $h = 1$.}
    \label{fig:schedule}
\end{figure}

A pause is where the value of $s$ is held fixed for a certain period of time, or equivalently the values of $A,B$ are held fixed. We use the notation $s_p$ to denote the value of $s$ at which a pause is inserted into the annealing schedule, and $t_p$ the time paused for, e.g. as shown in Fig.~\ref{fig:schedule}.
For our simulations, we fix the annealing schedule to be that of the D-Wave 2000Q device \cite{DW2KQ} that was used to collect experimental data (for more information on this device, see Ref.~\cite{powerOfPausing}).

\begin{figure}
    \centering
    \includegraphics[width=0.5\columnwidth]{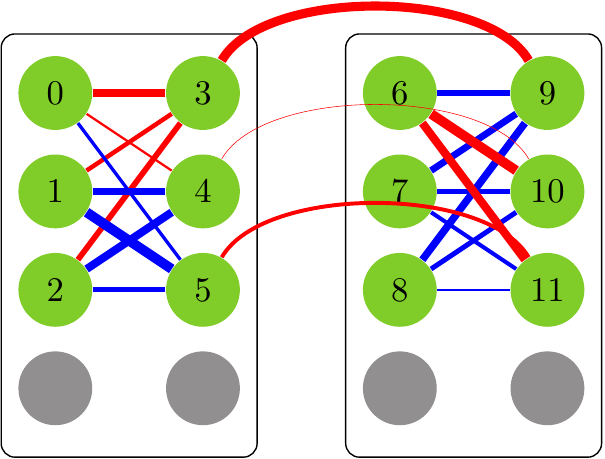}
    \caption{Depiction of instance $\mathcal{I}_{12}^0$ used in this study.  Shown are two Chimera unit cells, with the indexed nodes corresponding to the 12 qubits used in the instance.  Red (blue) edges connecting nodes correspond to ferromagnetic (antiferromagnetic) Ising couplings between qubits, with the thickness of the edge in direct proportion to the magnitude of the coupling.  Explicit values for the couplings are given in Appendix \ref{app:Instance}.}
    \label{fig:I12}
\end{figure}

\begin{figure}
    \centering
    \includegraphics[width=0.98\columnwidth]{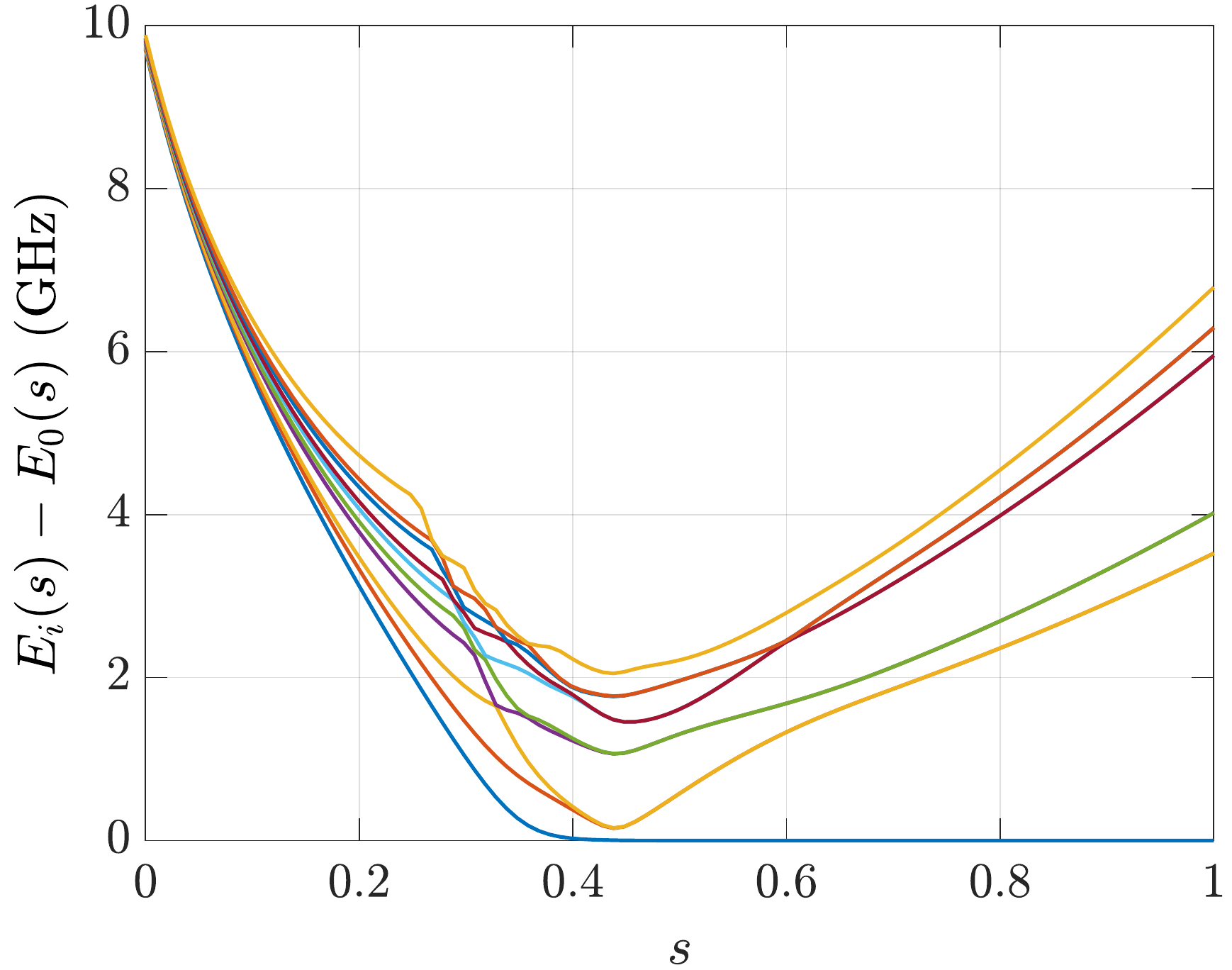}
    \caption{Spectrum (first 20 levels) of $\mathcal{I}_{12}^0$. Minimum gap of approximately $\Delta=0.15$ GHz at  $s_\Delta=0.44$. Specified DW operating temperature is 0.25 GHz. Units of $h=1$. } 
    \label{fig:spect}
\end{figure}

We focus our study on one well behaved instance $\mathcal{I}_{12}^0$ from Ref.~\cite{powerOfPausing}, which we depict in Fig.~\ref{fig:I12} (the instance is part of the supplemental material there and is given here in Appendix \ref{app:Instance}). It is small enough ($n=12$ qubits) to be amenable to analytic methods and simulations. This instance has no local-fields ($h_i=0$) thus has a doubly degenerate Ising spectrum, and the $J_{ij} \in [-1,1]$ are chosen according to a uniform distribution. Moreover, the minimum gap is less than the thermal energy scale (Fig.~\ref{fig:spect}), hence thermal effects are expected to play a role.

The minimum gap can be understood in terms of a perturbative crossing \cite{PhysRevA.80.062326}.  To see this, it is useful to work in the subspace with eigenvalue 1 under the operator $P = \prod_{i=1}^n \sigma^x_i$.  In the absence of dissipative dynamics, the unitary evolution would be restricted to this subspace if the initial state is the uniform superposition state.  In this subspace, the Ising ground state is given by:
\beq \label{eqt:GS}
\ket{E_0(s=1)} = \frac{1}{\sqrt{2}} \left( \ket{000110110000} + \ket{111001001111} \right) \ ,
\eeq
and we have two closely spaced ($\Delta_{21} \approx 0.0781$ in units of the Ising Hamiltonian) energy states above the ground state:
\bes \label{eqt:ES}
\begin{align}
\ket{E_1(s=1)} &= \frac{1}{\sqrt{2}} \left( \ket{001010001111} + \ket{110101110000} \right) \\
\ket{E_2(s=1)} &= \frac{1}{\sqrt{2}} \left( \ket{001110001111} +  \ket{110001110000}\right) \ ,
\end{align}
\ees
where the least significant bit in the state label corresponds to the state of qubit 1. These two energy eigenstates differ only in one position, i.e. are connected by the transverse field Hamiltonian $|E_1(s=1)\rangle = \sigma_9^x |E_2(s=1)\rangle$.
For a slightly different choice of $J_{ij}$'s, these two energy states would be degenerate, and our choice of $J_{ij}$'s weakly breaks this degeneracy.

\begin{figure}[htbp] 
   \centering
   \includegraphics[width=0.75\columnwidth]{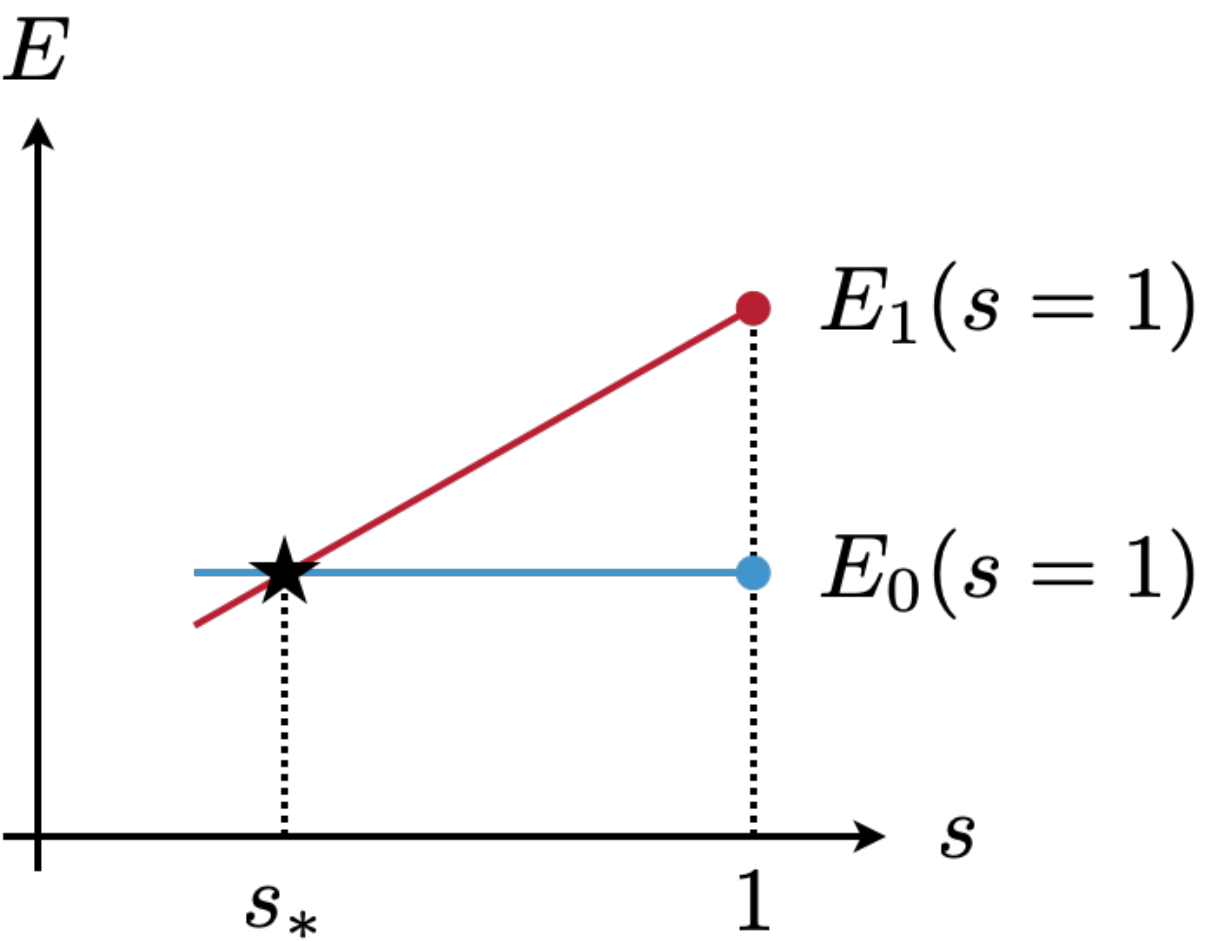} 
   \caption{Illustration of the energy level crossing predicted by first order perturbation theory.  At this order in perturbation theory, the ground state energy $E_0(s)$ remains unchanged while the first excited state energy $E_1(s)$ is lowered as a function of the perturbative parameter $1-s$, resulting in a energy level crossing at $s_\ast$.}
   \label{fig:crossing}
\end{figure}

Moving away from $s=1$ to smaller $s$ values, which corresponds to turning on the transverse field, first order perturbation theory predicts that the energy of the ground state $E_0(s)$ remains unchanged. In contrast, the symmetric combination of $\ket{E_1(s=1)}$ and $\ket{E_2(s=1)}$ is lowered in energy (the antisymmetric combination is raised in energy) at the same order in perturbation theory, and this combination is the unique first excited state away from $s=1$ with an energy $E_1(s)$ that decreases with decreasing $s$. 
First order perturbation theory thus predicts a crossing of the ground state and first excited state at some $s = s_\ast < 1$, as illustrated in Fig.~\ref{fig:crossing}, and 
this crossing is then corrected at higher order to an avoided level crossing, with the gap determined primarily by the Hamming distance between the computational basis states in Eqs.~\eqref{eqt:GS} and \eqref{eqt:ES} (at least 3 in this case). 

\section{Methods} \label{sec:Methods}
In this work we compare experimental results of the D-Wave 2000Q (DW) device \cite{DW2KQ} located at NASA Ames, to simulations of the adiabatic master equation (AME) and SVMC.

\subsection{AME} \label{sec:AME}

The AME is a time-dependent Davies master equation \cite{springerlink:10.1007/BF01011696,adiabaticME} of Lindblad form \cite{Lindblad:76} derived in the limit of weak coupling between the system and a Markovian environment.  We assume independent but identical Ohmic oscillator baths for each qubit, which gives rise to a spectral density of the form:
\beq
\gamma(\omega)  = \frac{2 \pi \kappa^2 \omega e^{- |\omega|/\omega_c}}{1 - e^{-\beta \omega}} \ ,
\eeq
where $\beta = 1/k_B T$ is the inverse thermal energy scale of the bath, $\kappa^2$ is the dimensionless system-bath coupling strength squared,  and we have introduced (by hand) the ultra-violet cutoff $\omega_c$.  Under these assumptions, the AME takes the form ($\hbar = 1$):
\begin{eqnarray}
\frac{1}{t_a} \frac{d}{ds} \rho(s) &=& -i \left[ {H(s)}, \rho(s) \right] + \sum_{i=1}^n \sum_{\omega} \gamma(\omega) \times \nonumber \\ 
&& \hspace{-1.75cm}  \left[ L_{\omega,i}(s) \rho(s) L^\dagger_{\omega,i}(s)  - \frac{1}{2} \left\{ L_{\omega,i}^{\dagger}(s) L_{\omega,i}(s), \rho(s) \right\} \right] \ ,
\end{eqnarray}
where the index $i$ runs over the $n$ qubits and the index $\omega$ runs over all possible Bohr frequencies of the system Hamiltonian $H(t)$.  The Lindblad operators are given by
\begin{eqnarray}
L_{\omega,i}(s) &=& \sum_{a,b} \delta_{\omega, E_b(s) - E_a(s)} \bra{E_a(s)} \sigma^z_i \ket{E_b(s)} \nonumber  \\
&& \times \ket{E_a(s)}\bra{E_b(s)} \ ,
\end{eqnarray}
where we have taken a dephasing system-bath interaction on each qubit and $\left\{ \ket{E_a(t)} \right\}_a$ are (instantaneous) energy eigenstates of $H(t)$ with eigenvalues $\left\{ E_a(t) \right\}_a$.

In the context of the AME, the increase in ground state probability associated with pausing after the minimum gap can be straightforwardly understood.  At any fixed $s$ value, the fixed point of the AME is the Gibbs state associated with $H(s)$.  As the gap opens up, the ground state population of the Gibbs state increases (ignoring pathological cases where maybe the first excited state degeneracy grows), and the dissipative dynamics restores population to the instantaneous ground state (thermal relaxation).  Whether taking an adiabatic (in the open system sense \cite{PhysRevA.93.032118}) anneal or pausing at a fixed $s$, the Gibbs state will be reached \cite{lorenzo-relaxation-adiabatic}; the relevant question though is how efficiently can this repopulation occur.  In the AME, this is determined by the non-zero eigenvalues of the Lindbladian.  These eigenvalues depend sensitively on the overlaps $\bra{E_a(s)} \sigma_i^z \ket{E_b(s)}$ and $\gamma(\omega)$, and the relevant first-excited state to ground state relaxation term in the Lindblad master equation is \cite{PhysRevA.91.062320}:
\begin{eqnarray}
\gamma_{1 \to 0} \propto \gamma(E_1(s) - E_0(s)) \sum_{i} | \bra{E_0(s)} \sigma_i^z \ket{E_1(s)}|^2 \ .
\end{eqnarray}
We further know that the ground state population in the Gibbs state increases to $1$ as $s \to 1$ (assuming the spectral gap is large here), while the overlap $\bra{E_a(s)} \sigma^z \ket{E_b(s)}$ goes to $0$.  Therefore, the presence of an optimal pause location after the minimum gap might be expected as a competition of these two effects.

\subsection{SVMC with transverse field updates \label{sect:svmc}}
As a brief reminder of SVMC~\cite{svmc}, each qubit is modelled as a two-dimensional rotor with an associated angle $\theta_i \in[0, \pi]$, where $\theta_i=0$ is aligned along the $z$-axis, and $\theta_i=\pi/2$ along the $x$-axis (here $i$ labels the qubit).

In the original model, update angles are chosen uniformly randomly, i.e. $\theta_i'$ has no relation to $\theta_i$. Updates to the spin angles $\theta_i \rightarrow \theta_i'$ are accepted according to standard Boltzmann factor associated with the change in energy of the classical Hamiltonian
\begin{eqnarray} \label{eqt:Hsvmc}
    \mathcal{H}(s) &=& -A(s) \sum_i \sin \theta_i  \nonumber \\ 
    && \hspace{-1cm} + B(s) \left(\sum_{i< j} J_{ij}\cos \theta_i \cos \theta_j + \sum_i h_i \cos \theta_i \right) \ .
\end{eqnarray}
This Hamiltonian can be interpreted as the semiclassical potential associated with the spin-coherent path integral \cite{PhysRevD.19.2349,Albash:2014if,Albash:16} collapsed to a plane \cite{PhysRevA.94.062106}.

This model fails to capture the phenomena of freeze-out that is present in current experimental quantum annealers. Freeze-out occurs in transverse field annealing with a dominant dephasing system-bath interaction when the transverse field is weak relative to the Ising Hamiltonian (i.e. `late' in the anneal where the instantaneous energy eigenstates are well approximated by the eigenstates of the Ising Hamiltonian), and the system effectively freezes with no more population dynamics occurring. In contrast, in SVMC where update angles are chosen arbitrarily, dynamics can still occur at $A = 0$. When angles are chosen near $0$ or $\pm \pi$, the updates correspond to  classical spin flips on the Ising Hamiltonian.  

Of course more general update procedures can be implemented, such as $\theta_i \rightarrow \theta_i + \epsilon_i(s)$, where $\epsilon_i(s)$ is associated with a random variable depending on $s$ (and $i$).  For example, in order to replicate the freeze-out effect, we consider updates where $\epsilon_i(s)$ is chosen randomly in the restricted range $ [-\min \left(1,\frac{A(s)}{B(s)}\right) \pi, \min \left(1,\frac{A(s)}{B(s)}\right) \pi]$.  For $A(s)/B(s) \gtrsim 1$ (corresponding to $s \lesssim 0.36$), the updates amount to the unrestricted updates of the original algorithm, but when $A(s)<B(s)$ the updates are restricted to be around the current angle value.
This simple adaptation induces freeze-out since now the transverse field strength directly determines how large an angle update is allowed, and in the limit of $A/B \rightarrow 0$, updates stop entirely.  To distinguish it from the standard SVMC, we refer to this new algorithm with transverse field dependent updates as SVMC-TF.

\begin{figure}[htbp] 
   \centering
   \includegraphics[width=0.95\columnwidth]{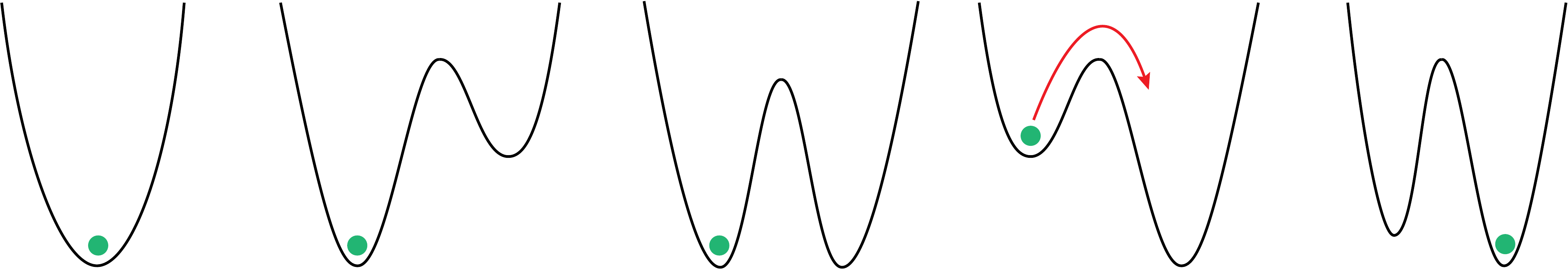} 
   \caption{Semiclassical picture of the dynamics.  Left (right) corresponds to early (late) in the anneal.  The green dot corresponds to the position of the system in the energy landscape.}
   \label{fig:SVMCPotential}
\end{figure}

The dynamics during the evolution of the classical energy landscape of SVMC and SVMC-TF (Eq.~\eqref{eqt:Hsvmc}) are illustrated in Fig.~\ref{fig:SVMCPotential}.  Early in the anneal, the semiclassical potential has a single minimum, and the dynamics  follows this minimum. As we approach the minimum gap, a second local minimum develops.  Near the minimum gap of the quantum Hamiltonian, there are two near degenerate minima in the semiclassical potential, and the only way for the system to reach the other minimum is via thermal hopping.  If this minimum is far away, corresponding to updating many spins by large angles, reaching this minimum via thermal hopping is suppressed.  Crucially, SVMC is not able to tunnel to the new minimum.  After the minimum gap of the quantum Hamiltonian, there is again a single global minimum in the semiclassical potential, and it is unlikely that a sizeable population has reached it.  Depending on the shape of the landscape, if the state is stuck in some local minimum, it may still reach the new global minimum in this regime via thermal hopping.

In SVMC and SVMC-TF , as an analogue for `time', the number of sweeps is used, where one sweep consists of checking whether each (classical) rotor in the model should be updated with a new orientation.
Pausing in SVMC and SVMC-TF , where the classical potential is fixed for a chosen number of rotor-updates, provides more opportunities for the state to escape local minima, and will generally result in a larger population in the ground-state.

\subsection{Simulation parameters}
Unless otherwise stated, in our AME and SVMC simulations, the temperature is fixed at 12mK, the purported temperature of the DW device used in our study.  In the AME simulations, we fix $\kappa^2 = 10^{-3}$ and $\omega_c = 8 \pi$ GHz.  We also restrict the dynamics to the lowest 16 energy eigenstates of the instantaneous Hamiltonian; we find that this is sufficient to maintain $\Tr \rho \approx 1$ numerically.

DW experiments and AME simulations are both performed with an annealing time of $1\mu$s. Each data point from DW is computed from (at least) $10^4$ samples (anneals), using (at least) 5 choices of gauges (for more information, see Ref.~\cite{powerOfPausing} where some of this data was originally presented).

In the SVMC and SVMC-TF simulations, we use a total of $10^4$ sweeps for the anneal unless otherwise noted, and we perform $10^4$ independent anneals per data point.

\section{Results}  \label{sec:Results}

In this section, we present our results comparing various properties of SVMC-TF, AME, and where possible, experimental results of the D-Wave annealing device. 
Note that the goal of this work is not to match explicit quantities through a choice of parameters. Indeed, there are many possible parameters available in each of the three systems. Rather our strategy is to make `reasonable' choices for the parameters, and study how the systems behave relative to one another. This allows us to see which dynamics are consistent in the different models, and those which may differ. When this is the case, we provide intuitive reasoning for it.

\subsection{Varying pause time}

\begin{figure}
    \centering
    \includegraphics[width=0.98\columnwidth]{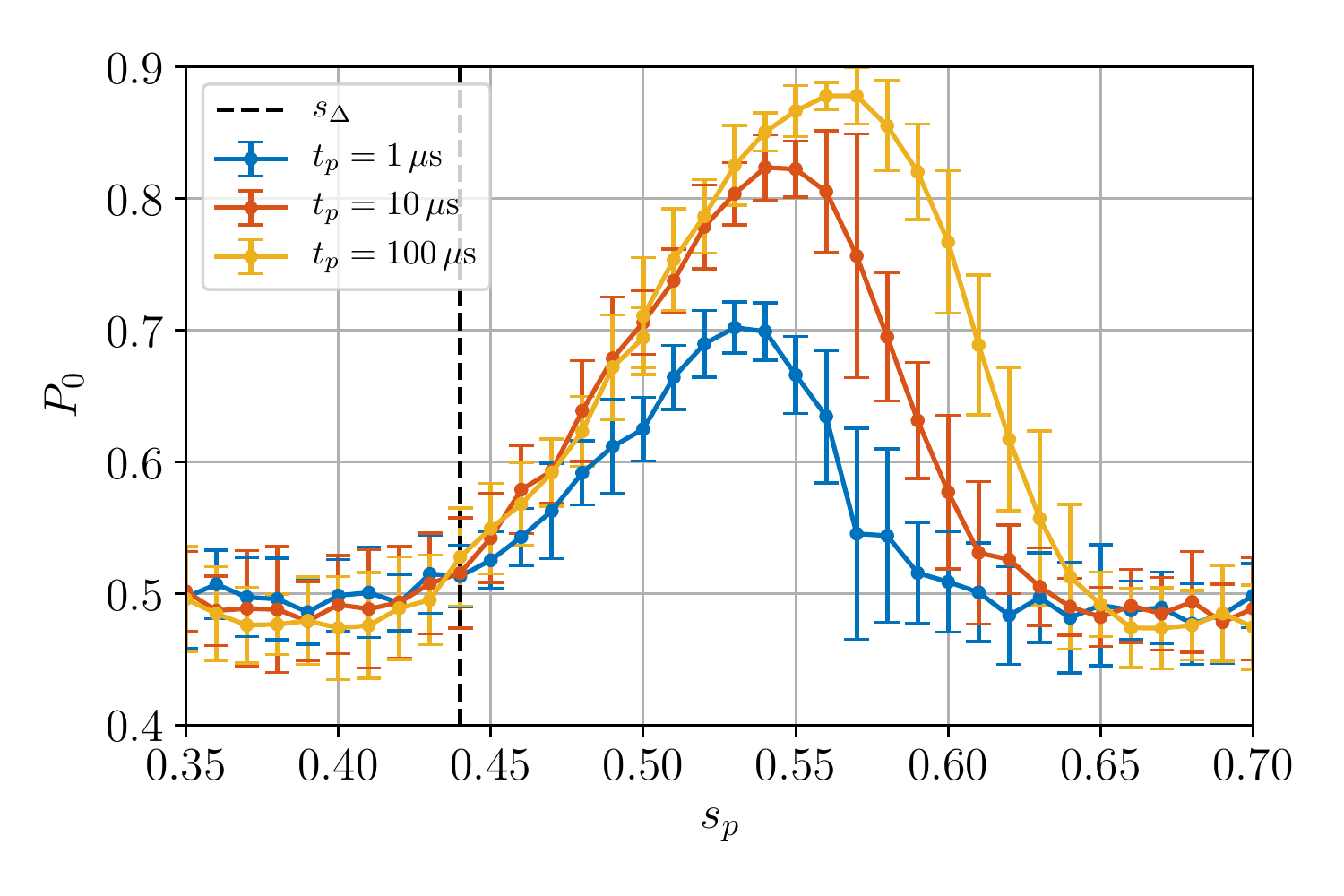}
    \includegraphics[width=0.98\columnwidth]{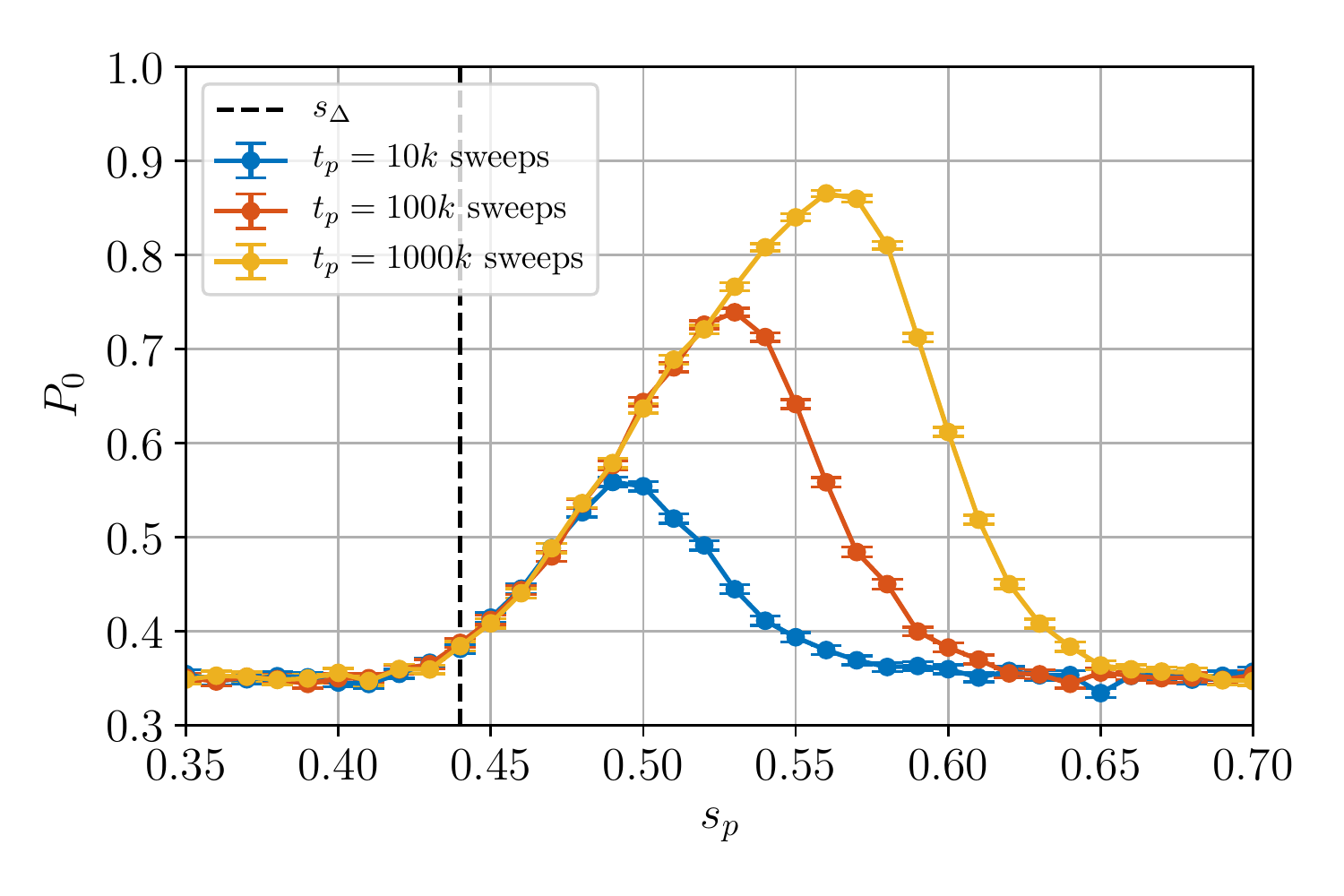}
    \includegraphics[width=0.98\columnwidth]{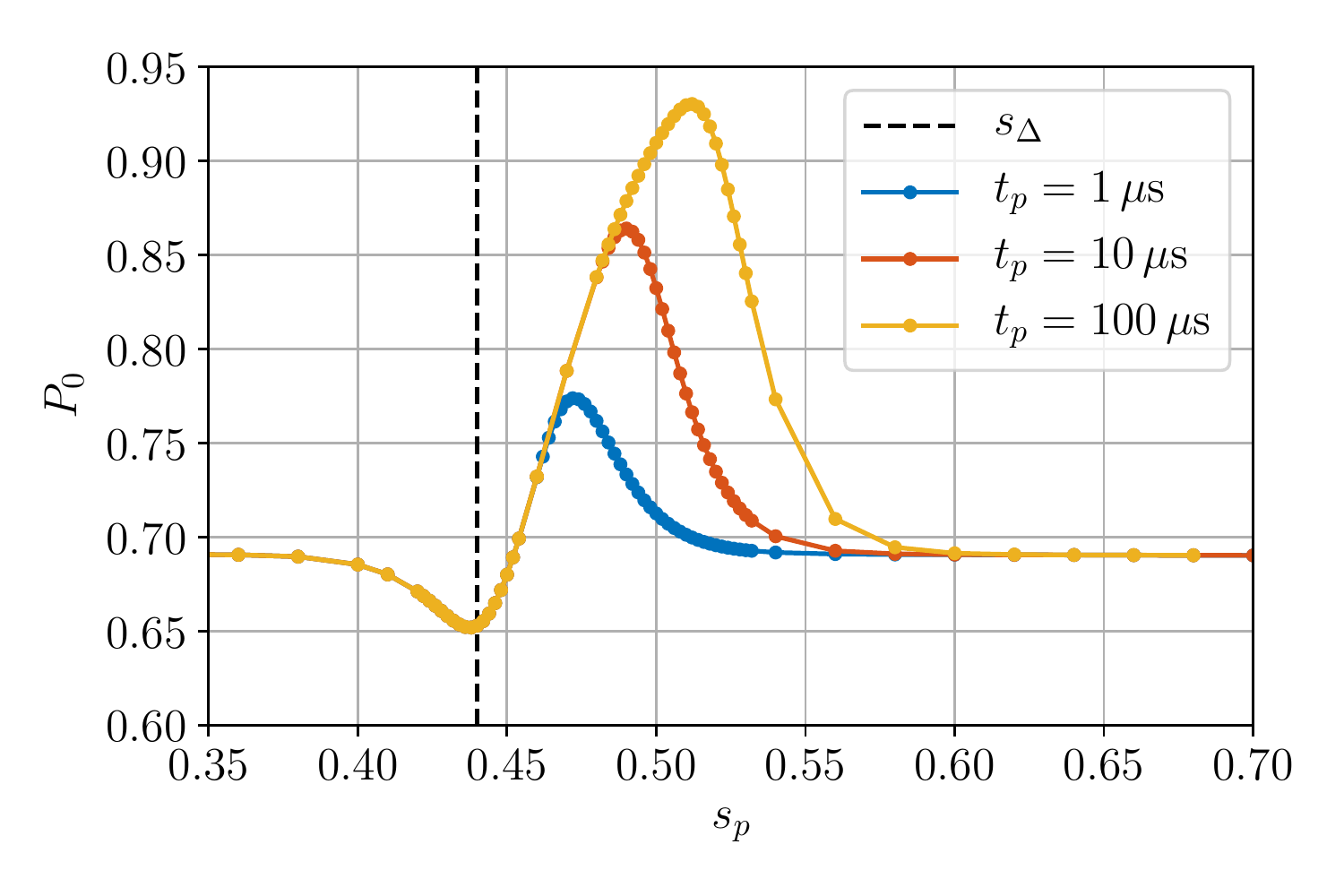}
    \caption{\textbf{Varying pause time for Top: DW, Middle: SVMC-TF, Bottom: AME}. Simulations all performed at 12mK. SVMC-TF has an anneal `time' of $10^4$ sweeps, and AME/DW of $1\mu$s. Note, for sake of comparison, the blue curves in each plot have a pause equal to the anneal time $t_a=t_p$, and the other two curves are 1 and 2 orders of magnitude larger.
    Error bars for DW are $2\sigma$ computed over at least 5 independent samples (each sample consisting of $10^4$ anneals).
    Error bars in SVMC-TF are $2\sigma$ over $10^4$ samples. {Note, the AME data is `exact'.} Vertical dashed line is position of the minimum gap $s_\Delta$. Notice the $y$-scale is different in all three plots.}
    \label{fig:vary_tp}
\end{figure}

As a first comparison, in Fig.~\ref{fig:vary_tp} we study the effect of changing the pause time, but keeping all other parameters fixed. The first clear observation is that both AME and SVMC-TF are broadly consistent with the experimental observation that increasing the pause time shifts the peak to a later time, and also increases its size. 

This shows that the mechanism by which the ground state re-populates under a pause is certainly not a uniquely quantum effect. Rather, it relies on there being an excess population in excited states at some late time in the anneal, and then, when pausing, for any type of relaxation process to take place. In SVMC-TF this corresponds to transitions from local minima in the potential at the pause location, to the global minima (see Fig.~\ref{fig:SVMCPotential}). In SVMC-TF, at a given pause point, and as we will explore in more detail below, the temperature is critical to driving these transitions, since no quantum tunneling can occur.
With the added restriction on the SVMC-TF updates, this adds a further constraint on how easy it is for SVMC-TF to transition to the global optima, at a given pause location.

Note that in all three figures there are differences as well as similarities. We stress here again that we are searching for broad properties, and not matching precisely the curves; there may well be sets of parameters in which all three models show closer agreement. 

Each model shows consistent behavior with respect to the position (and height) of the peak shifting later in the anneal with pause time $t_p$. This phenomenon is well understood and commented on in Ref.~\cite{powerOfPausing}, where a longer pause allows one to pause at a later time but still repopulate the ground state effectively (where otherwise the relaxation rate is reduced). The later the system can effectively thermalize, assuming the gap opens, the larger the instantaneous ground state (GS) population.

Another observation is that pausing at the minimum gap in DW and SVMC-TF does in fact provide a slight improvement in $P_0$, whereas in AME pausing at this point is detrimental. In AME, pausing at the minimum gap (the location where the Gibbs state has the lowest instantaneous GS population) means population is transferred to the excited states. If the dynamics after the anneal are too slow (e.g. ramp is too short), there is not enough time to effectively re-populate the GS. Indeed, we will see later that this dip disappears at higher temperature. (We note that for SVMC-TF, using a much smaller number of sweeps in the anneal can manifest a similar dip in $P_0$, but other features of the results get affected.  We present these extra results in Appendix \ref{sec:appendix-ramps}.) 

On the other hand, in the model of SVMC-TF pausing near the minimum gap allows for some of the population to explore the deep welled landscape and end up in what will eventually be the global minimum (cf. Fig.~\ref{fig:svmc_comparison}, middle). Based on the parameters of the model therefore (determining how easy it is to leave the current global minimum), pausing even at the minimum gap can eventually lead to larger population in the GS at $s=1$.  

The assumption of weak-coupling in the derivation of the AME is violated when the minimum gap is sufficiently small. Beyond weak-coupling, we can expect the broadening of the energy levels due to the interaction with the bath to be significant.  The dynamics of flux qubits in the strong bath coupling regime \cite{PhysRevA.94.062106} has important similarities to the dynamics of SVMC, so we can think of multi-qubit SVMC as a phenomenological model for a system of flux qubits in the strong coupling regime with no entanglement. The fact that DW exhibits behavior more akin to SVMC-TF than AME near the minimum gap is suggestive that we are observing a break-down of the weak-coupling limit in this region on the device. Various imperfections (control errors and noise) in the device are expected to broaden the spectrum which can modify the dynamics during this region \cite{noise-amp-annealing,googleTunnelingI}. In this case, the minimum gap may not be a well defined point, but more accurately modeled as a region, and this goes well beyond our AME model.

Finally we comment that in SVMC-TF, the point at which dynamics slow (here around $s_p=0.65$) is captured fairly well. As demonstrated in Appendix \ref{sec:appendix-svmc-comparision} (Fig.~\ref{fig:svmc_comparison}), the standard SVMC continues to exhibit dynamics for significantly longer $s_p$ values, inconsistent with experiments.

\subsubsection{Probing relaxation time \label{sec:relaxation-time}}
\begin{figure}
    \centering
    \includegraphics[width=0.98\columnwidth]{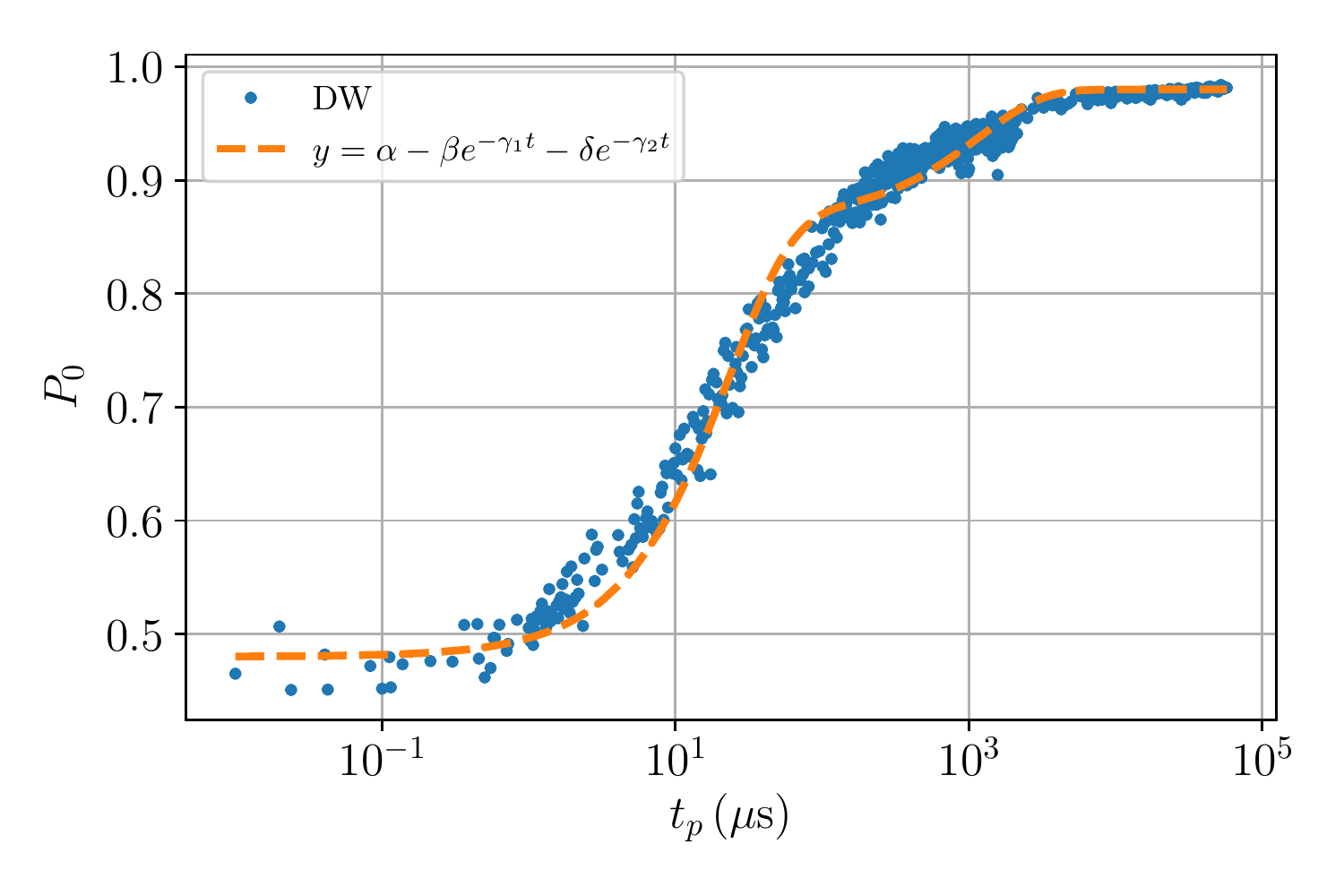}
    \includegraphics[width=0.98\columnwidth]{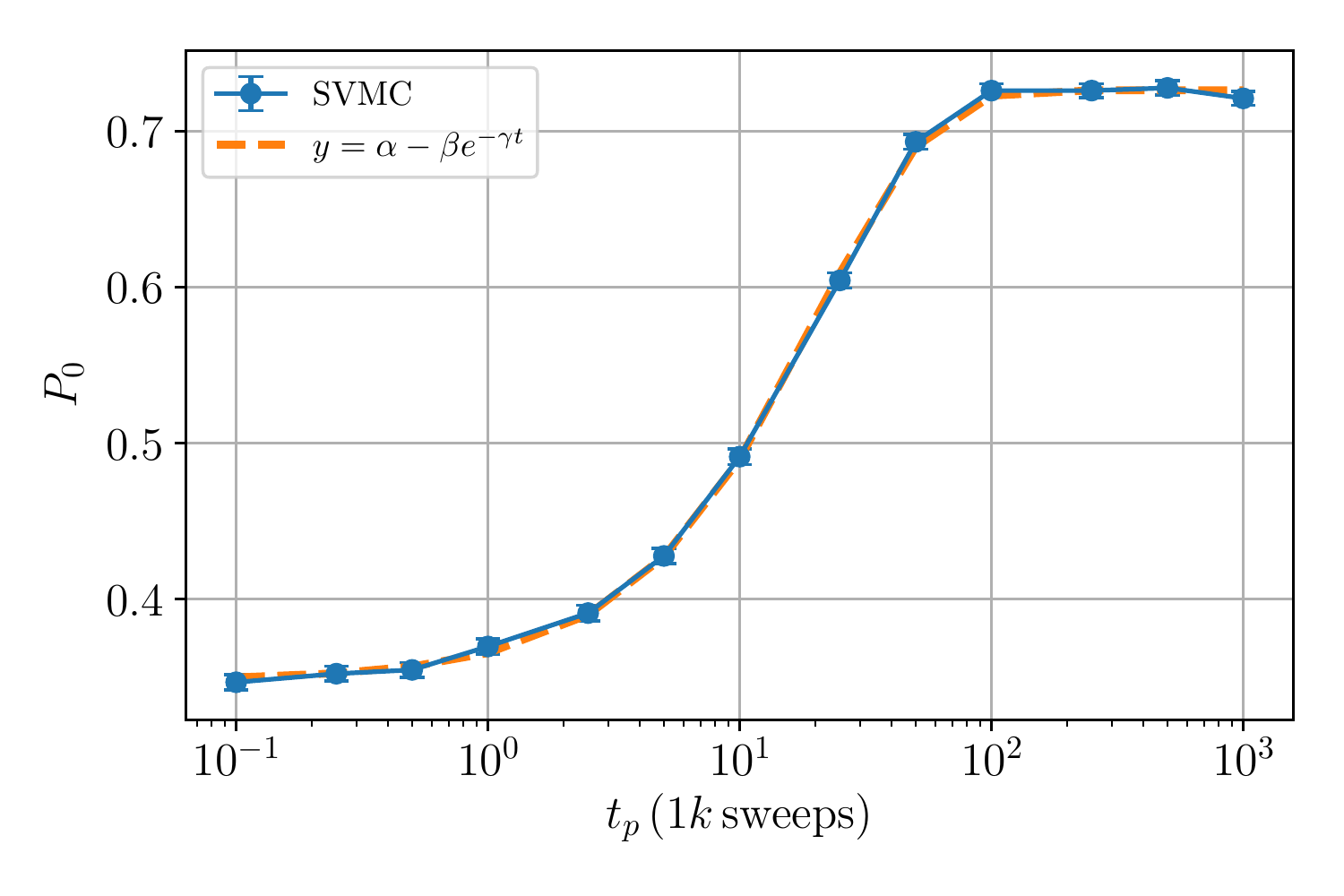}
    \includegraphics[width=0.98\columnwidth]{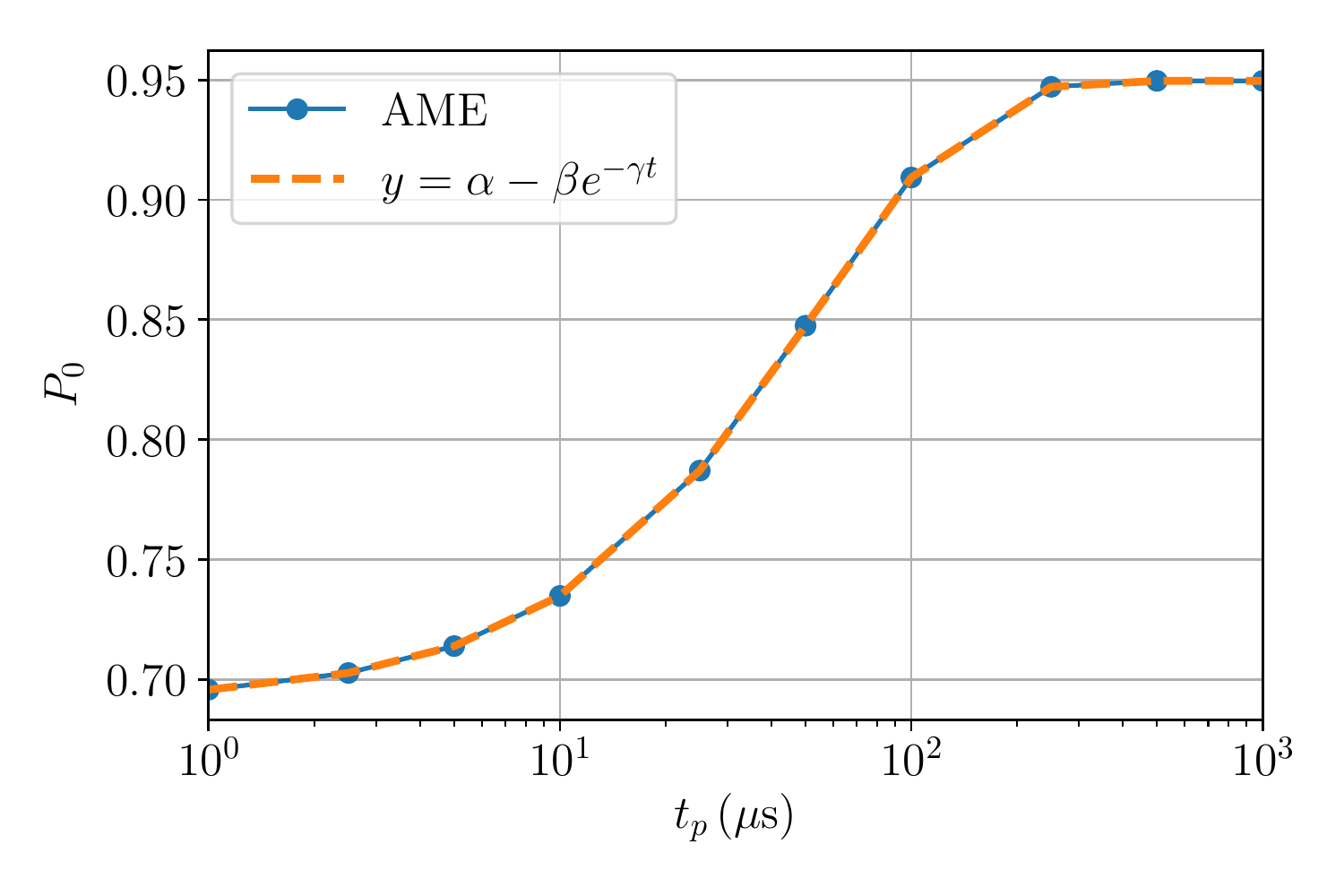}
    \caption{\textbf{Ground state probability with pause time for Top: DW, Middle: SVMC-TF, Bottom: AME}. {We fix the pause location in each plot to be in the region where the pause has a noticeable effect, taking $s_p=0.57$ for DW, and $s_p = 0.52$ for SVMC and AME}. The curve fits give excellent agreement with SMVC: $(\alpha,\beta,\gamma) = (0.73, 0.38, 0.047)$, AME: $(\alpha,\beta,\gamma) = (0.95, 0.26, 0.019)$. For DW, we had to use a function with two decay constants ($\gamma_1, \gamma_2) = (0.043, 8.5 \times 10^{-4})$. The top shows individual DW anneal results (from $10^4$ samples), and AME is `exact'. Error bars for SVMC are $2\sigma$ over $10^4$ samples.}
    \label{fig:P0_tp}
\end{figure}

The time required under the pause to reach saturation can be equated to the equilibration time. From experiments and simulations we can extract these characteristic times, which informs us about decay mechanisms involved.
Using the new extended pause time feature (allowing up to 60ms pause) we can start to see the approach to equilibrium on DW, which was previously unreachable (e.g. as seen in Fig.~6 of Ref.~\cite{powerOfPausing}). 
In Fig.~\ref{fig:P0_tp} we fix the pause location at the region around the peak observed in Fig.~\ref{fig:vary_tp}, and vary the time paused, while keeping all other parameters fixed.

Interestingly, at long pause times, the DW outperforms the AME simulations, which are (in principle) at the same temperature (DW reaching in excess of 95\% GS probability). Naively this may make one believe the DW is in fact operating at a colder temperature than the reported 12mK, although this would be contrary to several previous studies \cite{marshall-rieffel-hen-2017, powerOfPausing}, but the location of the peak for DW is at a larger $s^\ast$ value, which is consistent with a larger ground state population in the thermal state. One other possibility is that for increasing pause times the device is more susceptible to an effect known as spin-bath polarization, such that outcomes of successive anneals become correlated \cite{DW-spin-bath-polarization}. This is caused by the appearance of local fields which bias the spins alignments and can cause them to more easily align with previously found ground states, thus resulting in a larger population.

From Fig.~\ref{fig:P0_tp} we can extract the relevant time-scales for the three models.
We find that AME and SVMC-TF can be fit very well to a function depending on a single decay time-scale, which in AME corresponds to around 50$\mu$s. This suggests a single dominant decay channel is responsible for the repopulation effect.
In the D-Wave however, the fastest decay time-scale is around half this, 25$\mu$s, and there also exists a secondary much longer time-scale of the order 1200$\mu$s, although the curve fit here is not as accurate. Nevertheless, it is clear there is an additional longer time scale involved (cf. Fig.~\ref{fig:dw-single-time} in Appendix \ref{sec:appendix-dw-time-scale}). Since this only appears at times greater than around 1ms, this could be related to the spin-bath polarization effect.

\begin{figure}
    \centering
     \includegraphics[width=0.98\columnwidth]{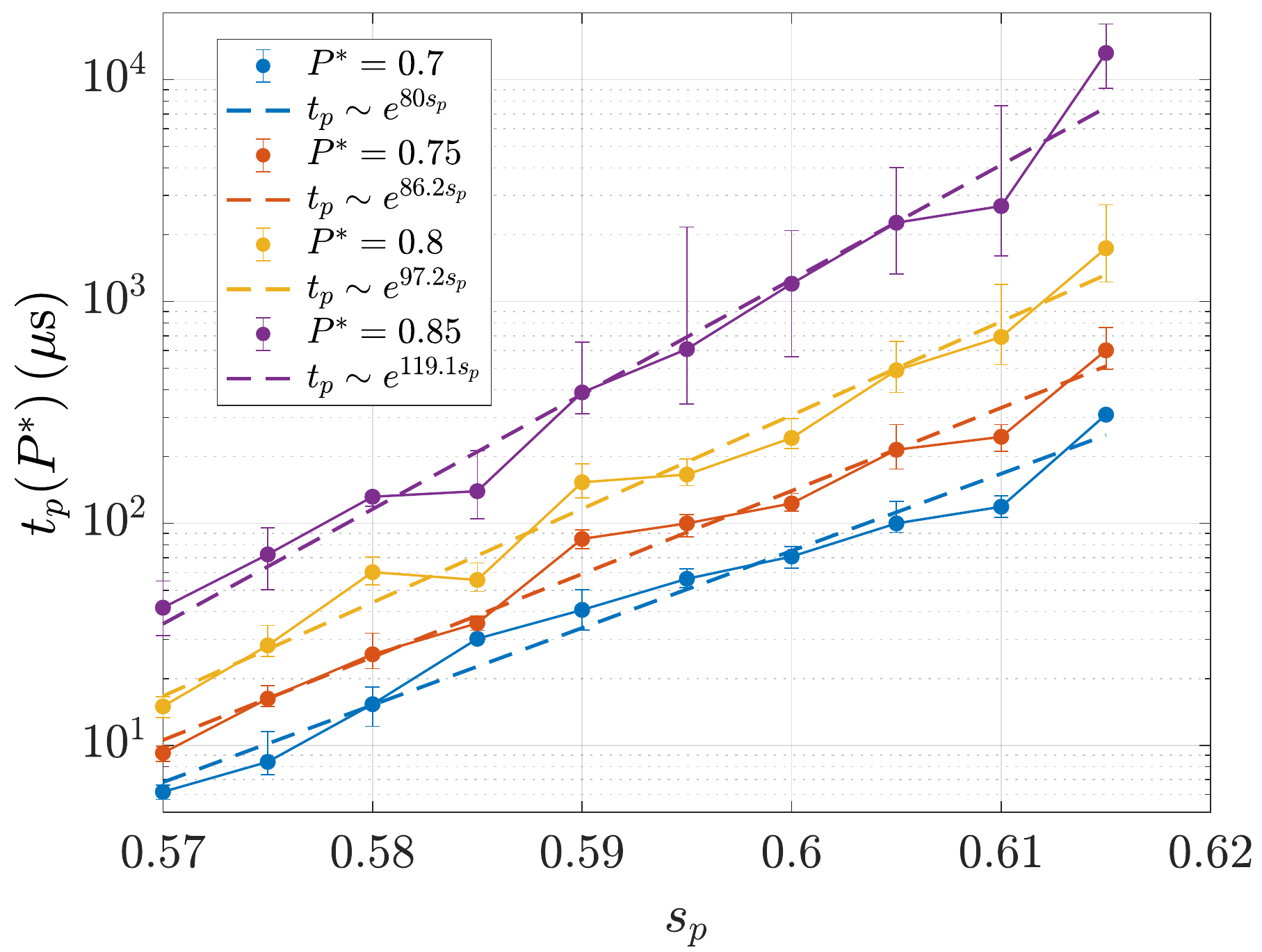}
    \includegraphics[width=0.98\columnwidth]{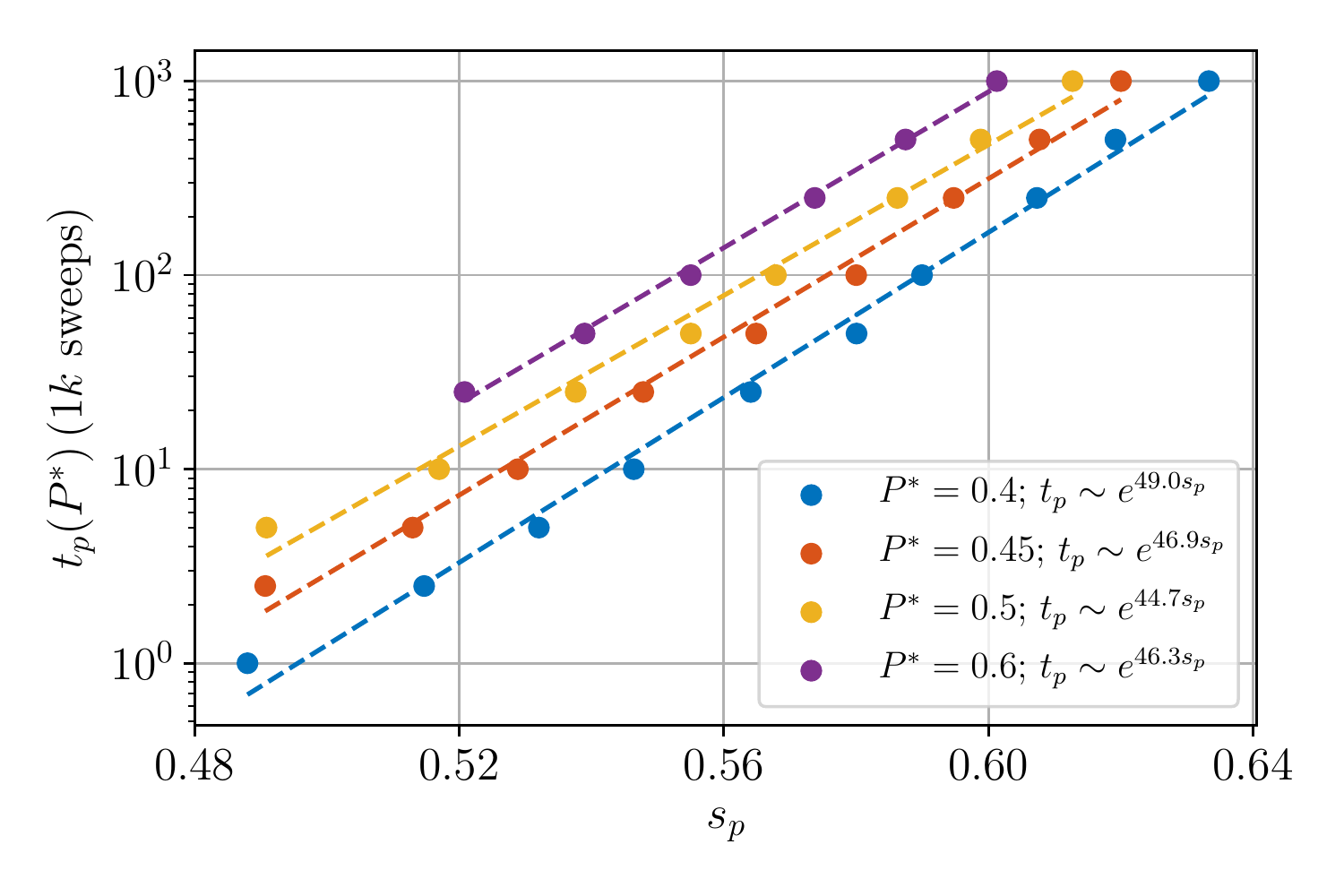}
    \includegraphics[width=0.98\columnwidth]{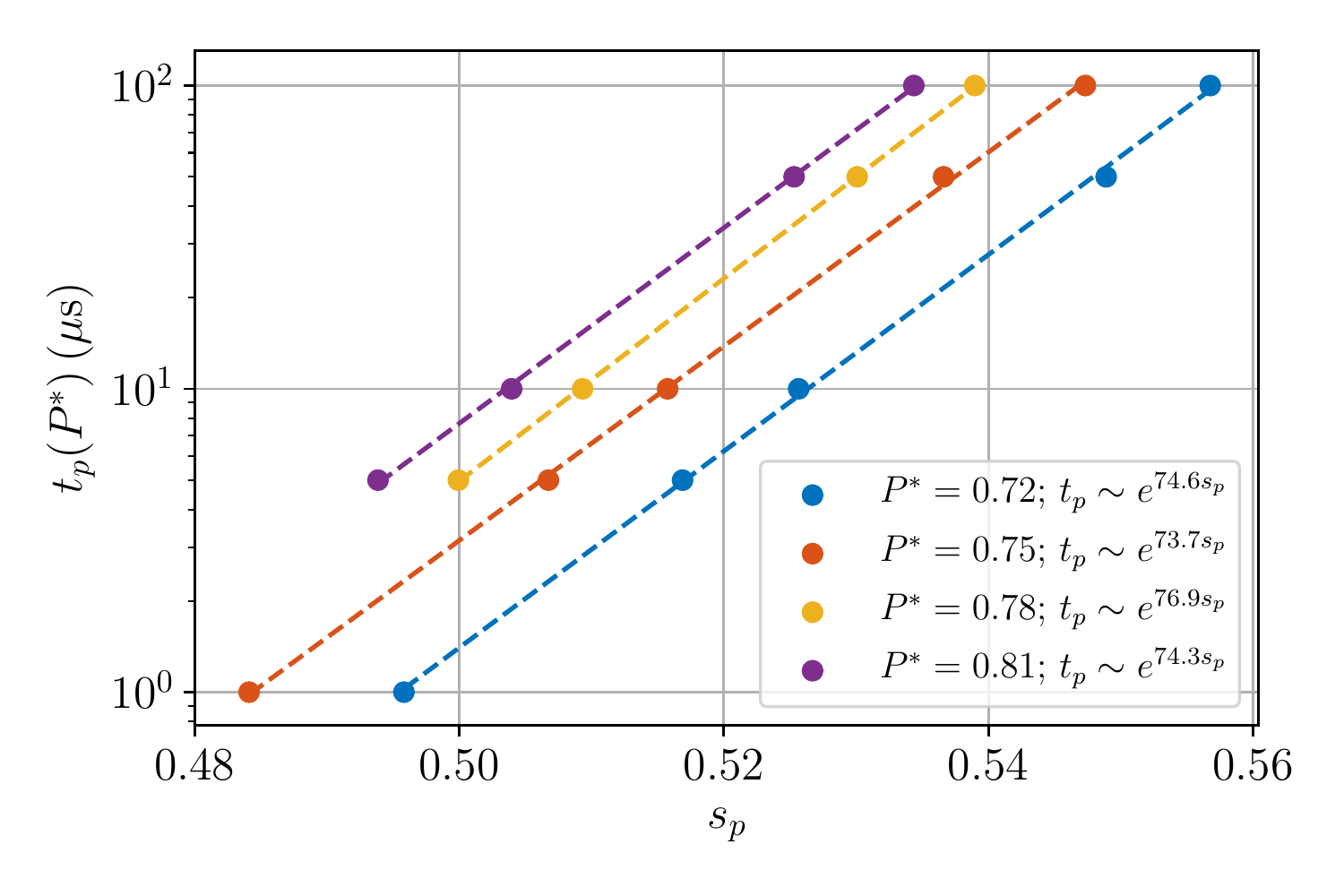}
    \caption{\textbf{Pause time required to reach target success probability $P^*$ for Top: DW, Middle: SVMC-TF, Bottom: AME.} The fits (dash lines) are from a least square fitting, with exponents given in the legend. Notice how the exponents change in the top plot compared to the bottom two. In AME and SVMC, the data is generated by taking curves as in Fig.~\ref{fig:vary_tp}, and finding the value on the $x$-axis corresponding to some fixed target $P^*$. If it lies between data points, we connect the points on either side by a straight line and extrapolate to find the value. {Since the raw DW data is noisier when pausing at a fixed location (e.g. as in Fig.~\ref{fig:P0_tp}), we do a similar analysis but take a range around $P^* \pm 0.01$}. For all data points in this range, the value reported here is the median, with error bars being the interquartile range.}
    \label{fig:tp_for_p0}
\end{figure}

We study the dependence of this time scale on the pause location, as shown in Fig.~\ref{fig:tp_for_p0}, where we plot the pause time $t_p$ required for the success probability to reach a target value $P^*$, pausing at different points $s_p$ (around the peak location in Fig.~\ref{fig:vary_tp}).

The results show a clear exponential increase in pause time required in order to obtain a particular ground state probability at the end of the anneal. This indicates that the relaxation rate is  decreasing exponentially in $s$ after the minimum gap, and is direct evidence that `freeze-out' occurs in all three models \cite{amin-freezeout, marshall-rieffel-hen-2017}, whereby the population dynamics slow and eventually stop late in the anneal.  To understand where this exponential dependence comes from in the AME, we note that  $\bra{E_1(s)} \sigma_i^z \ket{E_0(s)} \propto A(s)^\alpha$ for $A(s)/B(s)$ being small (see Appendix \ref{app:Aexp} for an argument), and since the annealing schedule $A(s)$ has an exponential form $A(s) \sim e^{-a s}$ (with $a\approx 30$ for $s$ approaching 1), it follows that the relaxation time scale in the AME will grow exponentially (see the discussion in Section~\ref{sec:AME}).  It is perhaps reasonable to expect SVMC-TF to replicate this given our definition of the updates, but we also find that standard SVMC exhibits similar behavior (shown in Appendix \ref{sec:appendix-svmc-comparision}, Fig.~\ref{fig:svmc_reg_pause_to_target}) indicating that the semiclassical landscape and the standard updates can still qualitatively capture this feature of the dissipative dynamics of the AME.

One interesting observation is how the exponents change with the probability target $P^*$. In experiments on the D-Wave (top in Fig.~\ref{fig:tp_for_p0}), the higher the target, the larger the exponent. This suggests it is relatively harder to reach a success probability of for example 90\% compared to 70\%. 
This is consistent with the observation of a slower decay channel, since it means it takes relatively longer to reach the highest success probabilities.
This is in contrast to the AME and SVMC-TF that exhibit approximately constant exponents across the various $P^*$, which is again consistent with the previous observation of a single relaxation time scale.

It can be argued that the open quantum system model of the AME or SVMC-TF is too simplistic to capture quantitatively the D-Wave output; this includes going beyond the weak coupling limit \cite{googleTunnelingI} but also unaccounted noise sources and effects, such as $1/f$ noise, which become more prevalent at the longer time scales required to reach high $P^\ast$, but also the fact that superconducting flux qubits only approximate a 2-level system.  If the temperature at which the device operates is in fact larger than documented (i.e. larger than 12mK, which has also been suggested previously \cite{powerOfPausing}), the population loss to higher excited states may be more severe than in the AME and SVMC-TF (at the minimum gap for example).

\subsection{Effect of temperature}
One important distinction between `quantum thermalization' and `classical thermalization'  is how the quantum system couples (e.g. through the $z$ component of the spin) to its thermal environment. The mechanism for thermalization then depends heavily on the transverse field of the system as ultimately this drives the transitions between computational basis states (cf. Fermi golden rule \cite{googleTunnelingI}). In particular, quantum systems can find their way out of local minima even at low temperatures when the barrier is higher than the energy scale set by the temperature. In the classical case, transitions of this type are exponentially suppressed.  Thus, we expect the temperature parameter to be critical to understanding differences between the two.

Although probing the temperature in experiments is not a current feature available to D-Wave users, we can gain an understanding through simulations.

\begin{figure}
    \centering
    \includegraphics[width=0.98\columnwidth]{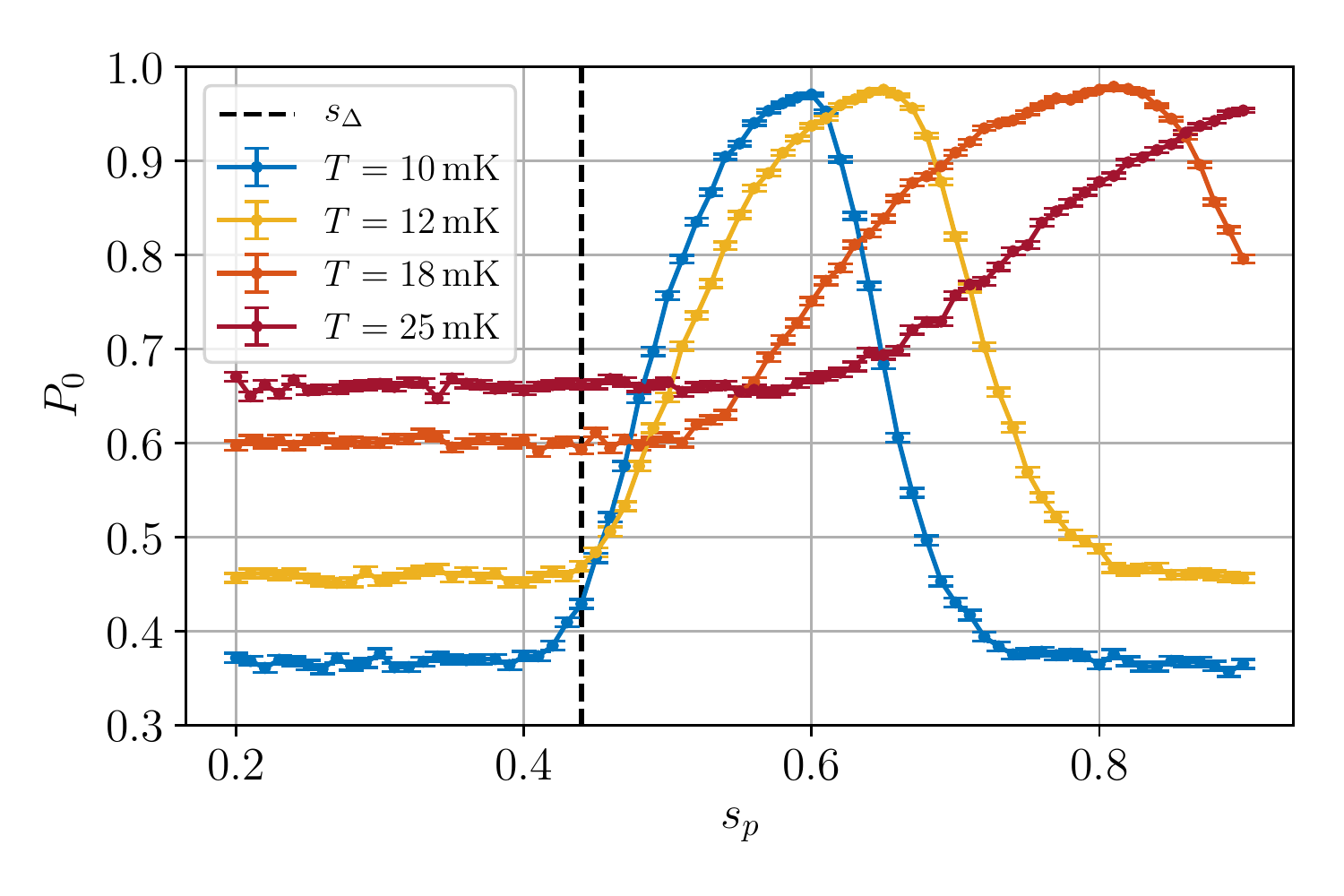}
    \caption{\textbf{SVMC dependence on temperature with random angle updates}. In the standard version of SVMC with new angles chosen uniformly randomly in $[0, \pi]$ dynamics continue late in the anneal, thus even at high temperatures a large success probability can be achieved, assuming the pause is late enough. The anneal `time' is $10^4$ sweeps, and the pause `time' is fixed at $10^6$ sweeps for all curves. Location of minimum gap $s_\Delta$ marked as vertical dashed line.}
    \label{fig:svmc-regular-update-temp}
\end{figure}

First, we note that the update procedure for SVMC is crucial here. As explained in Sect.~\ref{sect:svmc}, using the standard updates can lead to dynamics even at $s=1$. We see this clearly in Fig.~\ref{fig:svmc-regular-update-temp} where we show the success probability as a function of pause location for various temperatures. In SVMC the dynamics during a pause can be illustrated by considering the semi-classical potential, that changes as a function of $s$. At a higher temperature, it is required to pause later in order to observe the same increase in success probability, compared to a model at a lower temperature. The reason for this is that as $s\rightarrow 1$ valleys in the potential become deeper, and the hotter system can more easily jump out of the global minimum when the well is too shallow. Since the standard update procedure allows for one to in principle explore the entire landscape more readily, we do not see the characteristic freeze-out effect.

\begin{figure}
    \centering
    \includegraphics[width=0.98\columnwidth]{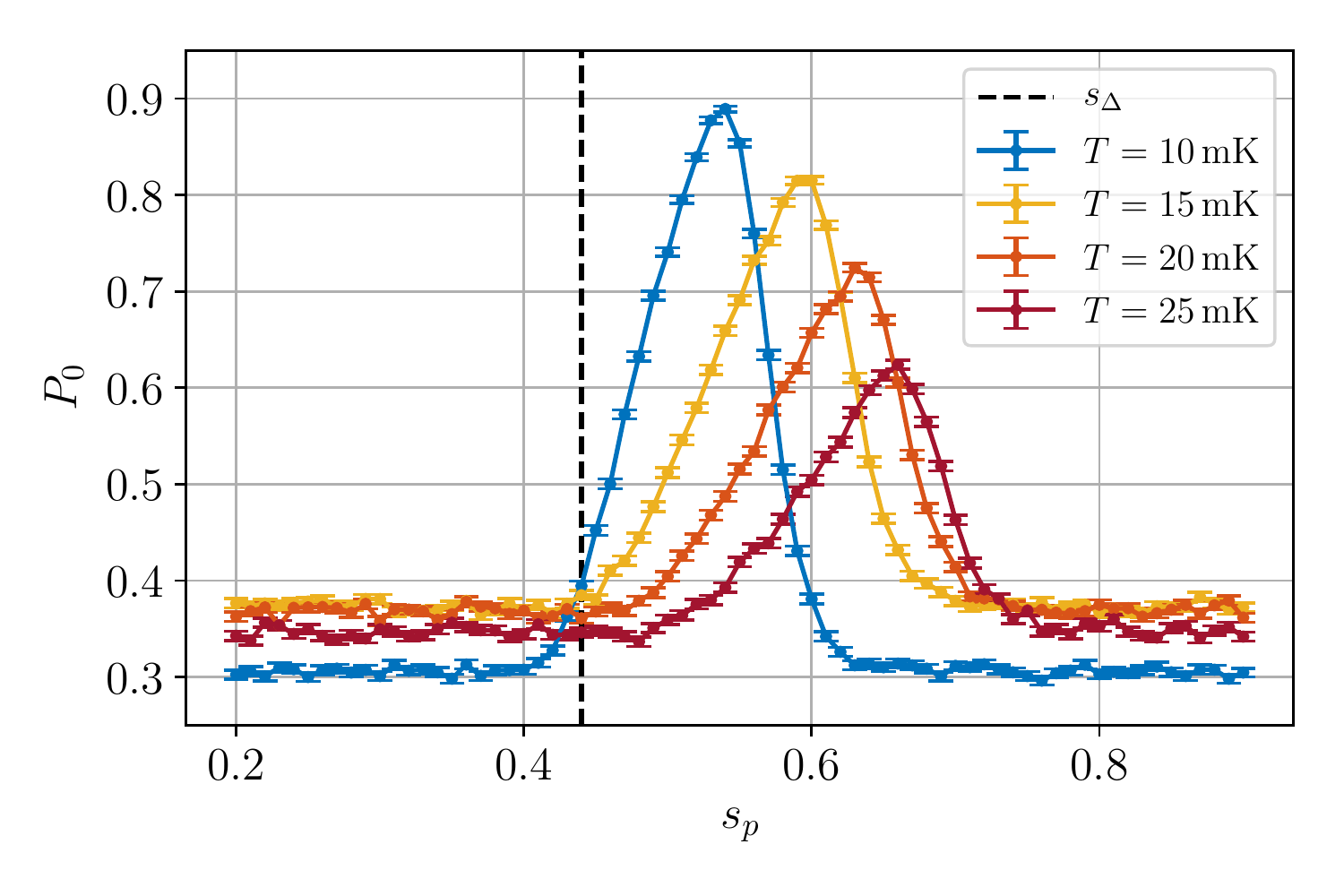}
    \includegraphics[width=0.98\columnwidth]{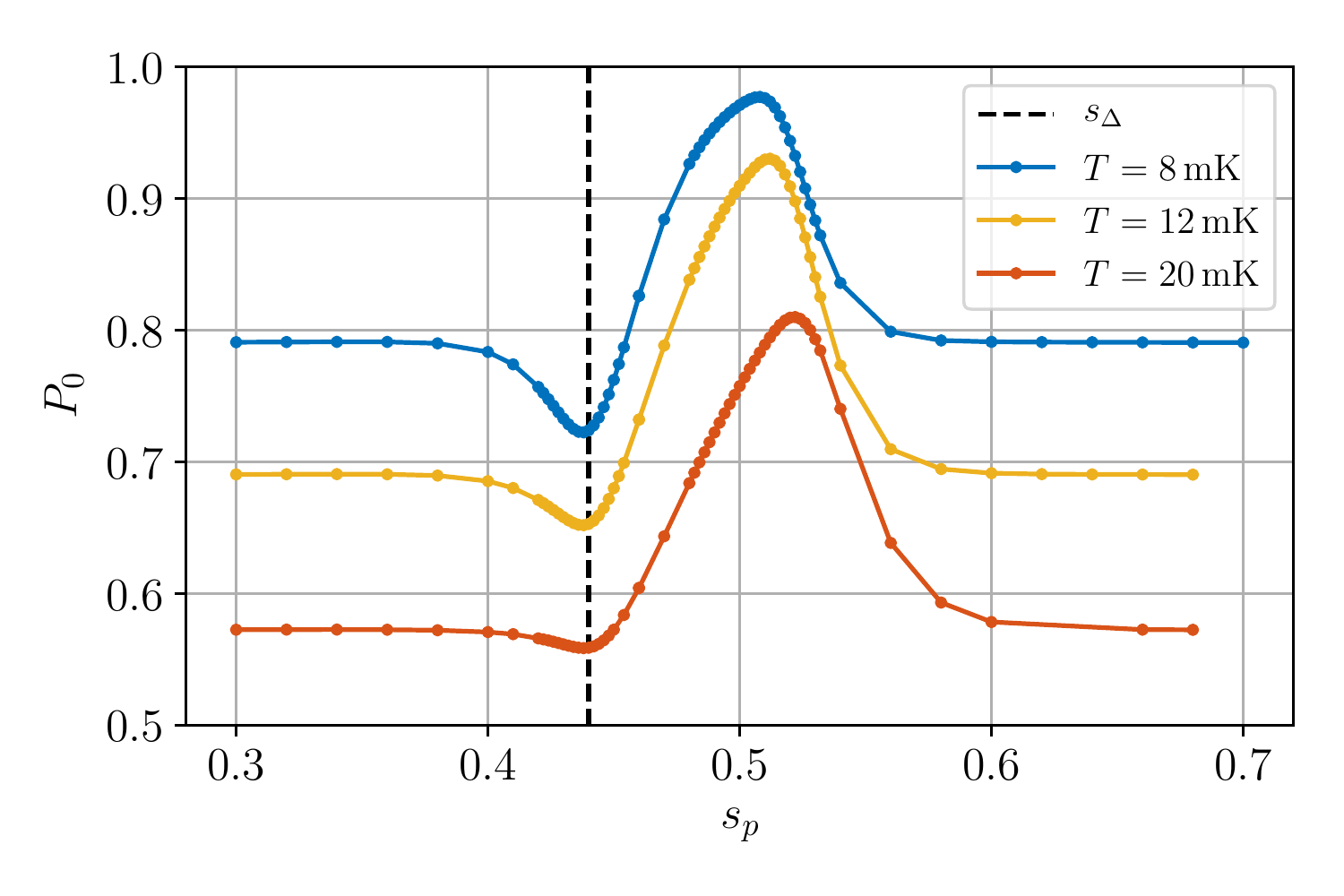}
    \caption{\textbf{Temperature dependence for Top: SVMC-TF, Bottom: AME}. Plots showing dependence of temperature under the pause at $s_p$. In SVMC, the anneal `time' is $10^4$ sweeps, and the pause `time' is fixed at $10^6$ sweeps for all curves. In AME we use an anneal of $1\mu$s and pause of $100\mu$s. Location of minimum gap $s_\Delta$ marked as vertical dashed line.}
    \label{fig:vary_temp}
\end{figure}

In contrast to this, Fig.~\ref{fig:vary_temp} (top) shows a dramatic change in the behavior of SVMC-TF. Here, more in line with intuition, the higher temperature system performs much worse compared to lower temperatures in the region where pausing has an effect. This is due to the freezing of the dynamics, since now the hotter system, which otherwise requires a pause late in the anneal to avoid jumping out of the global minima, has slowed dynamics in this region.

There are some intriguing differences between the dynamics of SVMC-TF compared to the AME however. In the AME, as expected from the general theory, a colder temperature is always best. This is because there are fewer thermal excitations out of the ground state in the first place since the Gibbs state has a higher GS population. As such, at a given pause location, a colder temperature always arrives at a greater population in the ground state. 
This is especially true for pausing both earlier in the anneal before the minimum gap and later in the anneal after the minimum, where we do not expect pausing to repopulate the ground state (early in the anneal the state already has overwhelming population in the ground state, while late in the anneal, thermal relaxation is slowed down considerably).  In both cases, the ground state population is determined primarily by how much population is lost due to thermal excitations by crossing the minimum gap (we note that in the closed system case, the ground state population would be almost 1 for these simulation parameters), which is reflected by their equal $P_0$ values for low and high $s_p$.

In SVMC-TF, while the peak GS probability increases with decreasing temperature, it is not generally true that a colder temperature results in a higher GS probability for a fixed pause location. 
We see this in Fig.~\ref{fig:vary_temp} (top), where e.g. when pausing after $s>0.6$, the hotter temperature can be the most effective. This is explained again by the semi-classical picture.  SVMC can not tunnel to the new global minimum but must reach it via thermal hopping. At a given pause location, a hotter temperature may enable the system to hop out of local minima more effectively than a colder system, and if a colder system pauses too late, it is likely to become stuck in sub-optimal configurations. 

We also mention that the behavior observed in the SVMC model is, of course, heavily dependent on the update scheme used. One can imagine different schemes where the angles are updated according to a different heuristic (i.e. other than used here with updates proportional to $A(s)/B(s)$). It may therefore be feasible that different schemes can give significantly different behaviour. We leave this for future study.

\section{Condition for advantage by relaxation}  \label{sec:Condition}
Based on the above observations of Sec.~\ref{sec:relaxation-time} where the dynamics are dependent on a single dominant decay rate, we provide a condition such that the time required to find the ground state is reduced. Note that in Ref.~\cite{chen-pause}, conditions for which an optimal pause location exists were derived under assumptions of the AME model, i.e., conditions for which there is a pause location which will increase the final ground state success probability. {Whilst the existence of an optimal pause location is a necessary prerequisite for reducing the time-to-solution (TTS) by pausing, this alone does not guarantee an improvement in performance, which must take into account the total time used to find a ground state (according to the TTS metric).}
The condition provided here is agnostic to the model by which thermalization occurs so long as there is a single dominant decay rate, and therefore applies to all three systems studied (DW, SVMC-TF, AME).

Assuming a functional form as empirically demonstrated in Fig.~\ref{fig:P0_tp}, {dependent on a single dominant decay rate $\gamma$ (in the DW case, neglecting the long time-scale, see Appendix \ref{sec:appendix-dw-time-scale})}, we can write the ground state probability measured as:
\begin{eqnarray}
\begin{split}
& P_0(s^*, t_a, t_p) = P_G(s^*, t_a)  \\ 
& - (P_G(s^*, t_a) - P_a(t_a))e^{-\gamma(s^*)t_p}
\end{split}
\end{eqnarray}
where $P_0(s^*, t_a, t_p)$ is the ground state probability at the end of an anneal with anneal time $t_a$ and a pause of length $t_p$ inserted at the optimal location $s^*$.
$P_a(t_a)$ is the ground state success probability achieved under a standard anneal with anneal time $t_a$ and no pause, and $P_G(s^*, t_a)$ the probability reached under an anneal with an \textit{infinite} pause at location $s^*$ (i.e. one reaching the Gibbs state), and with an anneal time $t_a$.  We expect $P_G(s^*, t_a)$ to have minimal, if any, dependence on $t_a$ because $s^\ast$ is typically after the minimum gap, and annealing from $s^\ast$ to $s=1$ should be effectively adiabatic.

The standard definition of the TTS is the time required to reach the ground state with 99\% probability:
\begin{eqnarray}
\mathrm{TTS}(s^*, t_a, t_p) = \frac{\log(1-0.99)}{\log(1 - P_0(s^*, t_a, t_p))}(t_a + t_p) \ .
\end{eqnarray}
Note that we are assuming the optimal pause location $s^*$ has already been identified.

\begin{figure}
    \centering
    \includegraphics[width=0.98\columnwidth]{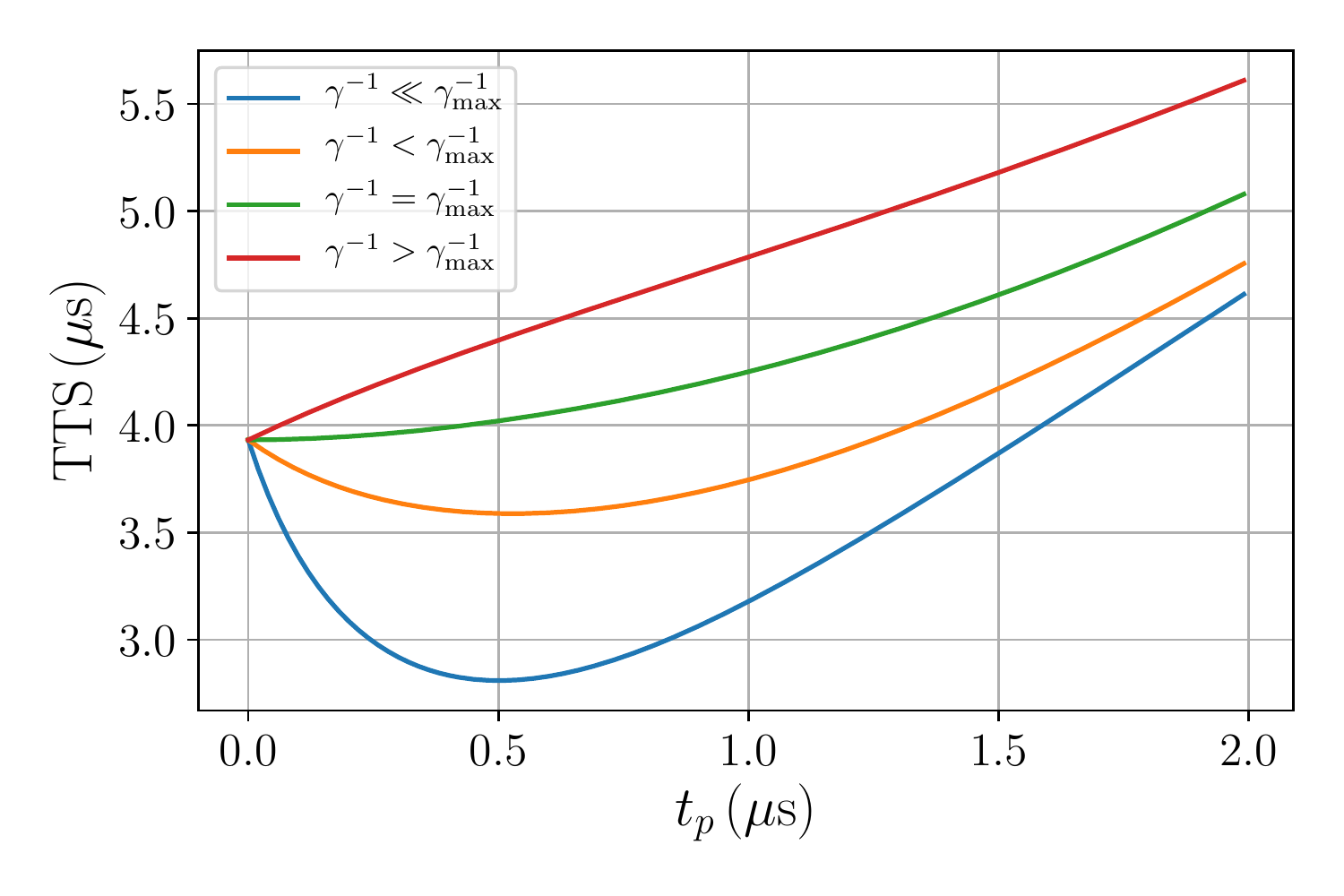}
    \caption{\textbf{Time-to-solution for various decay rates, as a function of pause time}. Here we plot using parameters of the AME as found in Fig.~\ref{fig:P0_tp} ($P_G=0.95, P_a=0.69, t_a=1\mu$s).  $\gamma^{-1}_{\mathrm{max}}$ is $0.72 \mu$s as shown in Table \ref{table:dtts}. When $\gamma^{-1}$ is small enough an initial decrease in TTS can be observed. For $t_p$ larger than shown, the curves eventually overlap and scale linearly in $t_p$.}
    \label{fig:tts}
\end{figure}

At long pause times, one expects TTS to increase as the approach to equilibrium occurs and any gain in success probability is over compensated by the increase in run time $(t_a+t_p)$, as shown in Fig.~\ref{fig:tts}.
With this, we assert that the TTS should initially decrease:
\begin{equation}
    \left. \frac{\partial \mathrm{TTS}}{\partial t_p}\right|_{t_p=0} < 0.
\end{equation}
This provides a condition,
\begin{eqnarray}
\label{eq:TTS-condition}
\gamma > \frac{-(1-P_a)\log (1-P_a)}{t_a (P_G - P_a)},
\end{eqnarray}
 stating that the decay rate (time) needs to be large (small) enough in order to observe a decrease in TTS. This condition on $\gamma$ is not simply about increasing the success probability; it reflects the \emph{rate} the success probability needs to increase whilst pausing, in order for it to be advantageous over a standard anneal. It gives a non-trivial relationship between the annealing dynamics that give rise to $P_a$ and the mostly static properties at $s^*$ that determine $P_G$. As an illustrative example, let us consider a scenario where $P_G = 1$ and $P_a = 1- e^{-c t_a}$, where $c$ denotes the convergence rate to the ground state due to annealing.  In this case, we find $\gamma > c$, indicating that in order for the pause to be beneficial, the thermal relaxation rate to the ground state must be faster than the annealing convergence rate to the ground state. 

 In Table \ref{table:dtts} we compute the maximum allowable decay time $\gamma^{-1}$ in order to observe a reduction in TTS, comparing to the parameters extracted from our simulations (Fig.~\ref{fig:P0_tp}). We see that in all three models, the measured value of $\gamma^{-1}$ is too large in order to observe a reduction in TTS. 
 
 In Fig.~\ref{fig:tts} we see that only for relatively short pause times can an improvement in TTS be observed. This is consistent with the results of Ref.~\cite{ferro-shift-pause} which found experimentally for a certain set of problems requiring embedding (problem size more than 100 qubits), that an improvement in TTS was observed for pause times less than around $2\mu$s.
 
 Given a decay rate $\gamma$ extracted from experiments/simulations, it is possible to find the optimal pause time by setting $\partial \mathrm{TTS}/\partial t_p =0$ (around $0.5 \mu$s in Fig.~\ref{fig:tts}).

\begin{table}
\centering
\begin{tabular}{ |c|c|c| } 
 \hline
 Model & $\gamma^{-1}$ for TTS reduction & $\gamma^{-1}$ measured \\ 
 \hline
 DW & $1.5\,\mu$s & $23 \,\mu\mathrm{s}$ \\ 
 SVMC-TF & $14k$ sweeps & $21k$ sweeps \\ 
 AME & $0.72 \,\mu$s & $54 \,\mu\mathrm{s}$ \\
 \hline
\end{tabular}
\caption{Table of values for the three models decay time $\gamma^{-1}$ (from Fig.~\ref{fig:P0_tp}), compared to the required value to observe a reduction in TTS. In all three models, $\gamma^{-1}$ is too large. }
\label{table:dtts}
\end{table}

\section{Conclusion}  \label{sec:Conclusions}

By performing a detailed analysis of a small instance with a sufficiently small gap, we have shown that the characteristic effect of improved ground state probability observed under pausing in transverse-field annealing is certainly not unique to quantum models.  Our model of open system dynamics has been the quantum AME, while our classical simulations have been based on the SVMC algorithm, which with simple modifications can be made to exhibit many of the thermal effects such as freeze-out observed with experiments on D-Wave.
We found significant qualitative agreements between all three results, which is perhaps striking given how SVMC and SVMC-TF do not have any detailed model of a thermal environment with which the system interacts.  This casts doubt on whether the origin of an improved success probability  whilst pausing is an inherently quantum one, though our analysis has been for a single problem instance so we caution against generalizations. There may be problem classes that exhibit drastically different behavior for these models and favor a quantum model under pausing.  We also should not discount the possibility that the physical process is significantly faster than a simulation of the process.  For example, one can view the results of Ref.~\cite{King2019} as a positive indication of physical thermal processes exhibiting a faster time scale than the simulated process. In our analysis, we have focused on the probability of measuring the ground state of the Ising Hamiltonian, which is the relevant task for solving classical optimization problems, so we believe that identifying classical optimization problems that can take advantage of these faster physical thermal processes remains an important research direction.  

The three models studied here all exhibit a single dominant decay channel causing the re-population of the ground state, which we used to derive a simple condition (Eq.~\eqref{eq:TTS-condition}) for which pausing can lead to a reduction in the time-to-solution. The temperature features prominently in this condition, since it determines the Gibbs ground state probability $P_G$ and the default annealing probability $P_a$ (from Eq.~\eqref{eq:TTS-condition}), but it also appears to be the most relevant tool to elucidate the dynamics involved in non-zero temperature annealing. We have seen that the two models studied here behave qualitatively differently as a function of temperature, which can serve as a useful benchmark for experimental devices \cite{googleTunnelingI}.
 
The role of temperature is expected to become even more important at larger system sizes. Since we know the Gibbs ground state probability decreases exponentially with increasing system size at a fixed temperature \cite{tempScalingLaw}, in order to maintain an advantage acquired with pausing with increasing system size will require the temperature to also decrease with system size.  With increasing system size, it seems reasonable to expect $\gamma^{-1}$ to scale exponentially for problems with exponentially small gaps \cite{lorenzo-relaxation-adiabatic}, so it is likely that the parameter regimes under which the TTS may be improved are not practically scalable.  This conclusion is not surprising given that we should only expect scalable quantum advantages in the context of fault-tolerant quantum computing.  Nevertheless, we believe our analysis will be helpful going forward in determining whether or not a pause can be beneficial for quantum annealing devices of a fixed size.

\begin{acknowledgments}
Computation for the work described in this paper was supported by the University of Southern California's Center for High-Performance Computing (hpc.usc.edu) and by ARO grant number W911NF1810227.
The research is based upon work (partially) supported by the Office of
the Director of National Intelligence (ODNI), Intelligence Advanced
Research Projects Activity (IARPA) and the Defense Advanced Research Projects Agency (DARPA), via the U.S. Army Research Office
contract W911NF-17-C-0050. The views and conclusions contained herein are
those of the authors and should not be interpreted as necessarily
representing the official policies or endorsements, either expressed or
implied, of the ODNI, IARPA, DARPA, or the U.S. Government. The U.S. Government
is authorized to reproduce and distribute reprints for Governmental
purposes notwithstanding any copyright annotation thereon.

JM is grateful for support from NASA Ames Research Center, the AFRL Information Directorate under grant F4HBKC4162G001, the Office of the Director of National Intelligence (ODNI) and the Intelligence Advanced Research Projects Activity (IARPA), via IAA 145483, and NASA Academic Mission Services, Contract No. NNA16BD14C.

\end{acknowledgments}

\bibliography{refs.bib}

\newpage
\appendix
\section{Instance $\mathcal{I}_{12}^0$} \label{app:Instance}
We give here the Ising parameters of instance $\mathcal{I}_{12}^0$ used in the main text.
\begin{eqnarray}
J_{0,	3} &=& -0.888765722269 \nonumber \\
J_{1,	3} &=&	-0.453396499878\nonumber \\
J_{2,	3}&=&	-0.581810391599\nonumber \\
J_{0,	4}&=&	-0.222181654366\nonumber \\
J_{1,	4}&=&	0.623744373452\nonumber \\
J_{2,	4}&=&	0.805987681935\nonumber \\
J_{0,	5}&=&	0.333955924275\nonumber \\
J_{1,	5}&=&	0.995412322296\nonumber \\
J_{2,	5}&=&	0.490983144977\nonumber \\
J_{3,	9}&=&	-0.925420427917\nonumber \\
J_{6,	9}&=&	0.663343935819\nonumber \\
J_{7,	9}&=&	0.687446523051\nonumber \\
J_{8,	9}&=&	0.749085209325\nonumber \\
J_{4,10}&=&	-0.0559945397502\nonumber \\
J_{6,	10}&=&	-0.990358090729\nonumber \\
J_{7,	10}&=&	0.491802375676\nonumber \\
J_{8,	10}&=&	0.505416377921\nonumber \\
J_{5,	11}&=&	-0.400367703995\nonumber \\
J_{6,	11}&=&	-0.831748994702\nonumber \\
J_{7,	11}&=&	0.413887841297\nonumber \\
J_{8,	11}&=&	0.204601421856
\end{eqnarray}
\section{Argument for why $\bra{E_1(s)} Z \ket{E_0(s)} \propto A(s)^\alpha$} \label{app:Aexp}
We restrict our analysis to the subspace defined by having +1 eigenvalue under the operator $P = \prod_{i=1}^n \sigma^x_i$.  We define our Hamiltonian as $H(s) = B(s) \left( H_p + \frac{A(s)}{B(s)} H_x \right)$, where we will treat $A(s)/B(s) \equiv \lambda(s)$ as our perturbative parameter.  We use $\ket{E_i^{(0)}} \equiv \ket{E_i(s=1)}$ for $i=0,1$ as defined in Eqs.~\eqref{eqt:GS} and \eqref{eqt:ES}, as our zero-th order state in perturbation theory that we are expanding from, and we note that these states are eigenstates of the $Z=\sum_i \sigma_i^z$ operator.  If we consider an expansion of the form (unnormalized):
\beq
\ket{E_i(\lambda)} = \sum_{k=0} \lambda^k \ket{E_i^{(k)}} \ ,
\eeq
then the first non-zero term that contributes in the perturbative expansion is at fourth order in $\lambda$ (i.e. the Hamming distance between GS and first excited basis states at $s=1$):
\begin{eqnarray}
\bra{E_1(\lambda)} Z \ket{E_0(\lambda)} &=& \lambda^4 \left( \bra{E_1^{(0)}} Z \ket{E_0^{(4)}} \right.  \nonumber \\
&& \hspace{-3cm} \left. + \bra{E_1^{(1)}} Z \ket{E_0^{(3)}} +\bra{E_1^{(2)}} Z \ket{E_0^{(2)} } \right. \nonumber \\
&& \hspace{-3cm}  \left. + \bra{E_1^{(3}} Z \ket{E_0^{(1)}}  + \bra{E_1^{(4)}} Z \ket{E_0^{(0)}} \right) + O(\lambda^6) \ .
\end{eqnarray}
Therefore, we can expect that $\bra{E_1(s)} Z \ket{E_0(s)} \propto A(s)^\alpha$.

\section{Role of the Annealing Time \label{sec:appendix-ramps}}
\begin{figure}
    \centering
    \includegraphics[width=0.98\columnwidth]{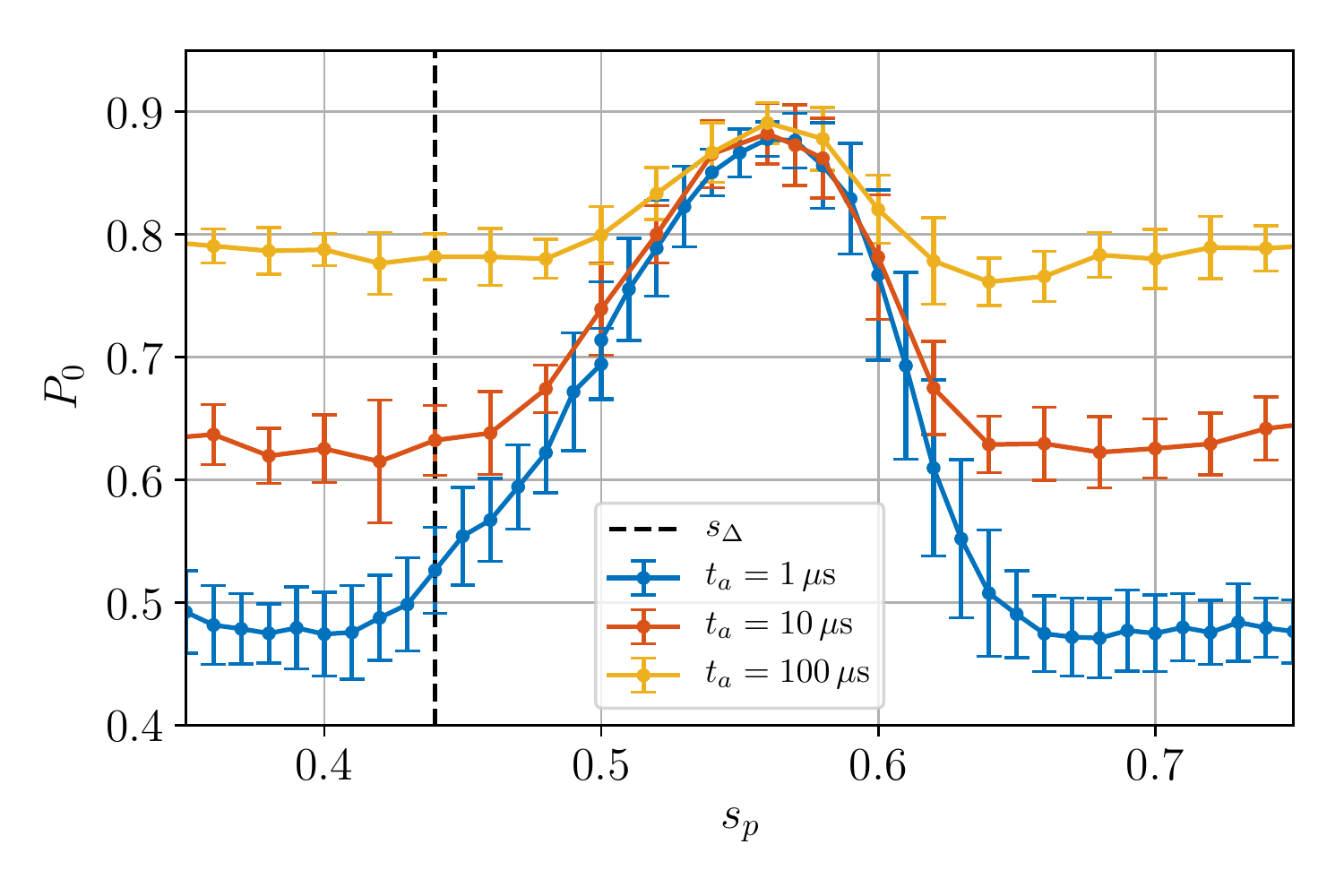}
    \includegraphics[width=0.98\columnwidth]{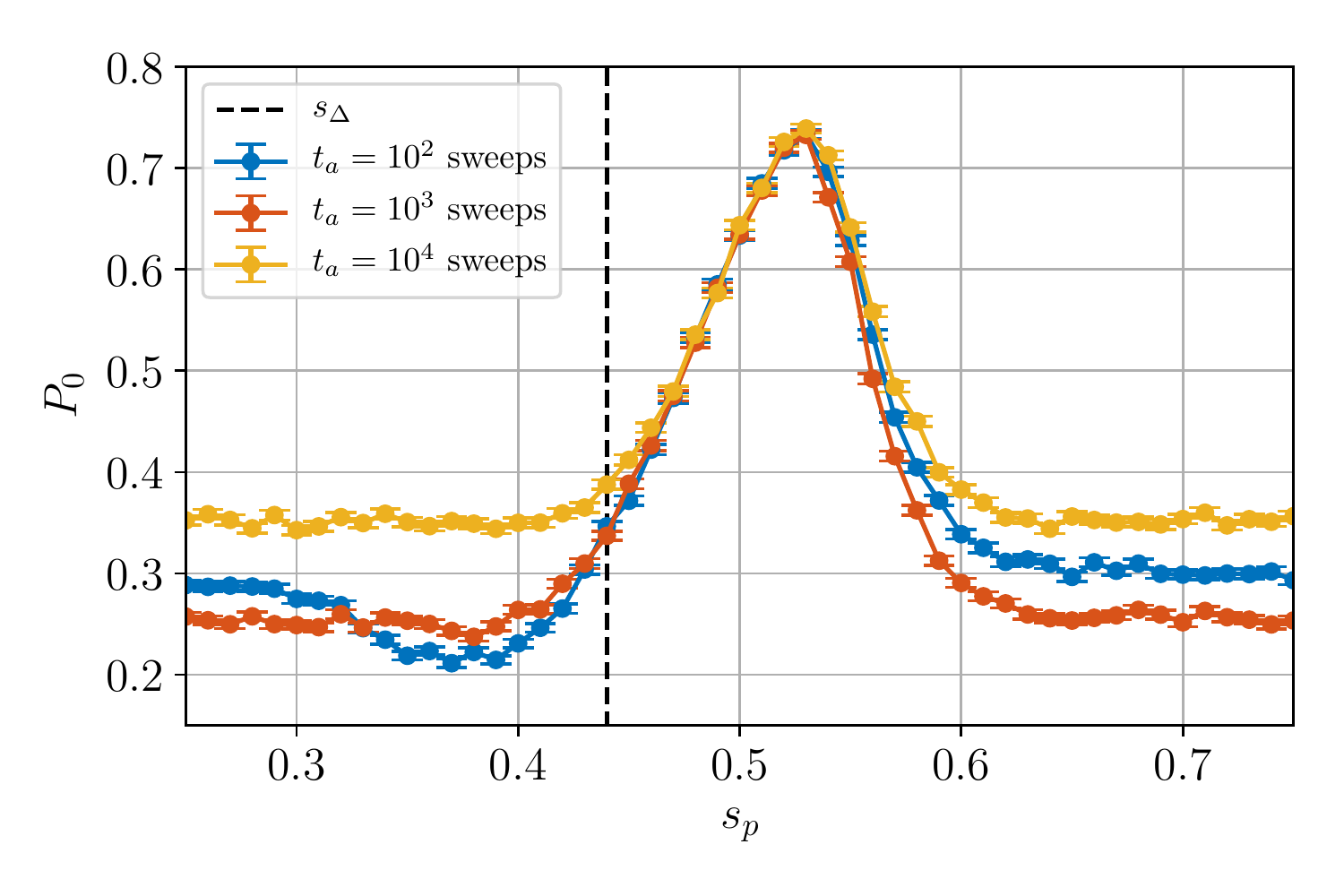}
    \caption{\textbf{Comparison of different annealing rates.} Ground state probability reached by DW (top) and the SVMC-TF algorithm (bottom) for a fixed pause duration $100\mu$s for DW, and $100k$ sweeps of SVMC-TF, but a varied annealing rate.}
    \label{fig:svmc_different_anneal}
\end{figure}

For the results in the main text, we kept the annealing rate fixed.  We show in Fig.~\ref{fig:svmc_different_anneal} how the annealing rate can change the qualitative features of the DW and SVMC-TF results.  For DW, increasing the annealing rate increases the GS probability for the early and late $s_p$ values but does not change the location of the optimal pause point or its peak value.  This is  similar to the behavior observed for SVMC-TF.

For SVMC-TF with very few annealing updates, we observe a similar drop in GS probability as for the AME (see Fig.~\ref{fig:vary_tp}), because with fewer sweeps during the anneal process, SVMC-TF is less likely to escape the local minimum and reach the global minimum after the pause.  However, at these small annealing rates, we begin to observe differences in the GS probability for early and late pause locations.

\section{Comparing SVMC to SVMC-TF \label{sec:appendix-svmc-comparision}}
\begin{figure}
    \centering
    \includegraphics[width=0.98\columnwidth]{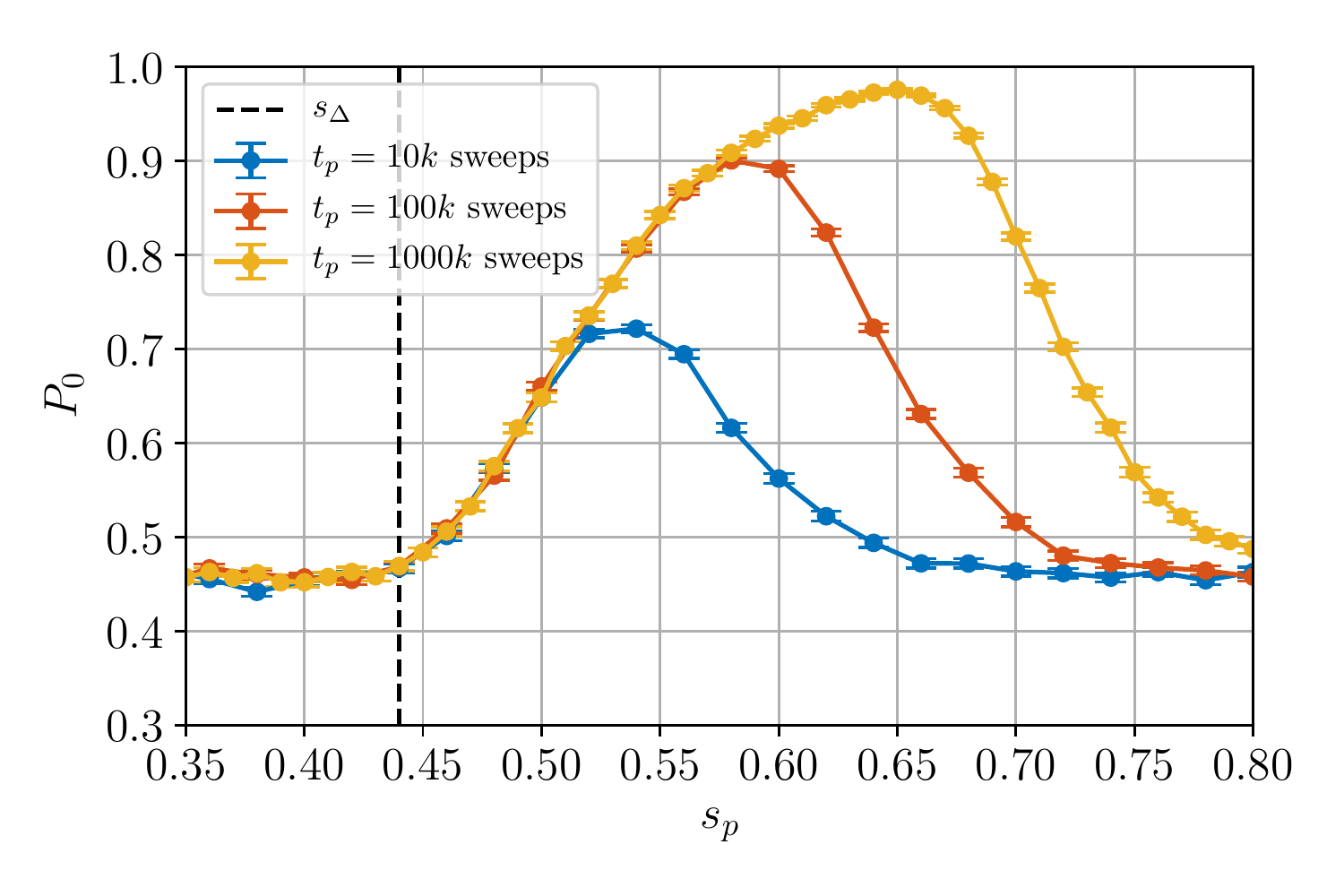}
    \includegraphics[width=0.98\columnwidth]{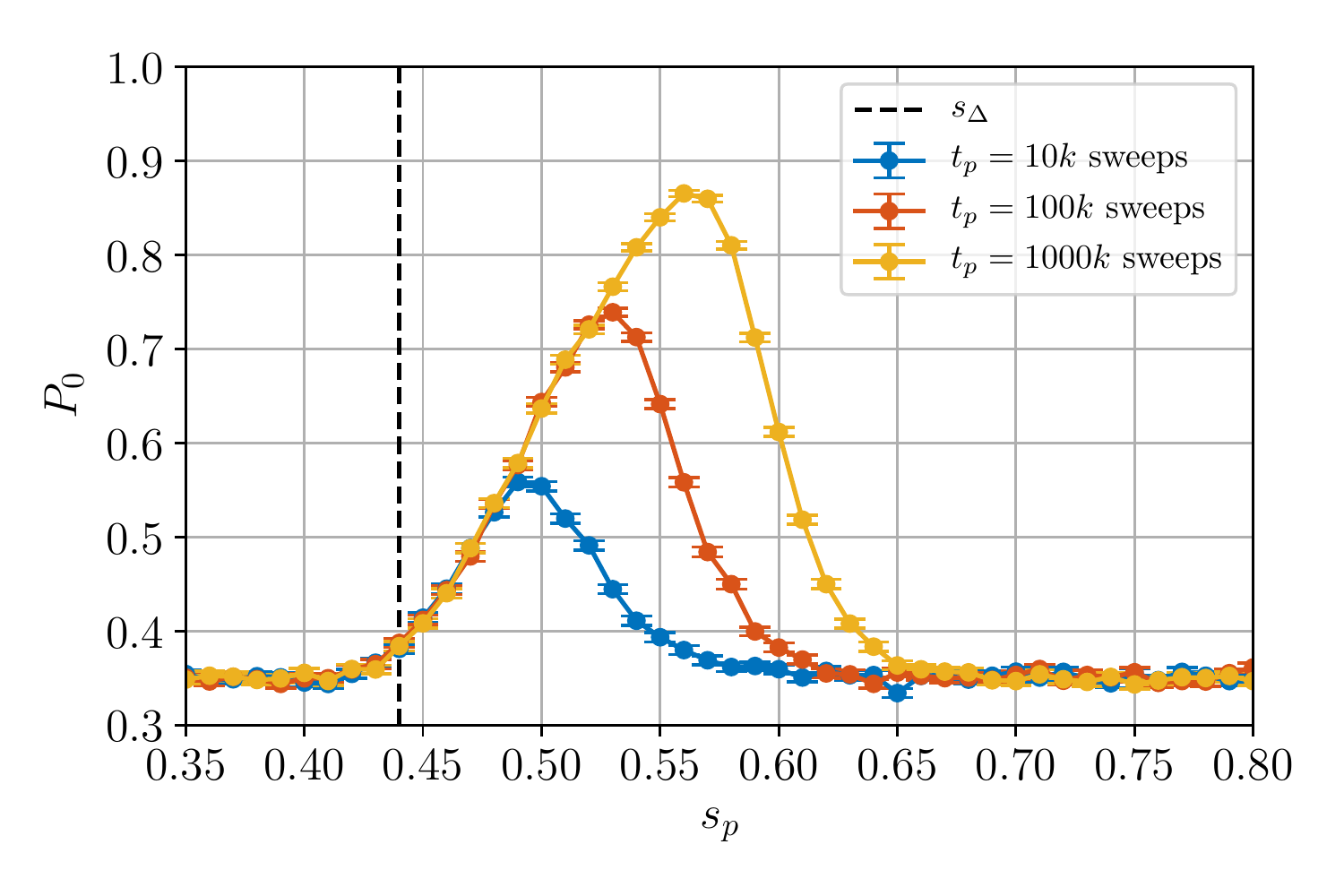}
    \caption{\textbf{Comparison of standard SVMC (top) with transverse-field updates SVMC (bottom)}. These simulations are both performed at 12mK with 10$k$ sweeps in the ramp. Notice the standard SVMC continues to have updates much later in the anneal (n.b. at $s=0.75$, the ratio $A/B < 10^{-2}$). Here we use an anneal `time' of $10^4$ sweeps for all curves.}
    \label{fig:svmc_comparison}
\end{figure}

Here we provide some additional comparisons between SVMC and SVMC-TF.  We show in Fig.~\ref{fig:svmc_comparison} how the two differ in terms of changing the number of sweeps while keeping the temperature fixed.  We see that SVMC does not exhibit as pronounced a suppression in the ground state probability for large $s_p$ as SVMC-TF does, which can directly be attributed to the modification in rotor updates we discuss in the main text.

\begin{figure}
    \centering
    \includegraphics[width=0.98\columnwidth]{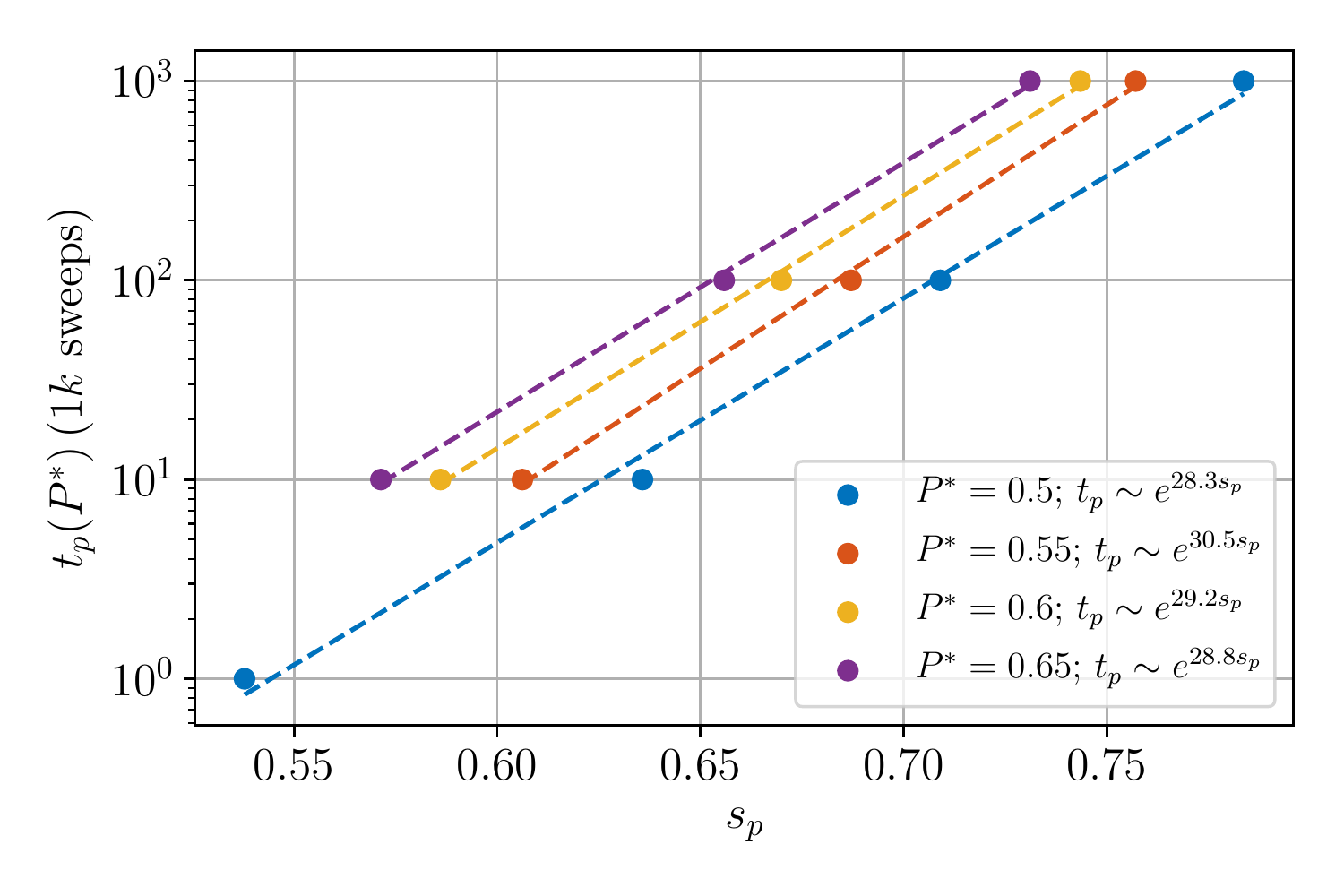}
    \caption{\textbf{Standard update SVMC pause time to target.} Pause time required for the standard SVMC algorithm to reach target success probability $P^*$. As in SVMC-TF the time required increases exponentially in $s_p$ (location of the pause).}
    \label{fig:svmc_reg_pause_to_target}
\end{figure}

In Fig.~\ref{fig:svmc_reg_pause_to_target} we show the pause time required for the standard SVMC algorithm to reach a target success probability (cf. Fig.~\ref{fig:tp_for_p0} in the main text). As in SVMC-TF, we see here an exponential increase in the time required to pause to reach the target success probability, though the scaling is more mild compared to SVMC-TF, due to the fact the dynamics slow much more in SVMC-TF late in the anneal. Moreover, the dynamics persist much later in the standard SVMC (as seen by comparing the scales on the $x-$axis).

\section{Relaxation time-scales in DW \label{sec:appendix-dw-time-scale}}
\begin{figure}
    \centering
    \includegraphics[width=0.98\columnwidth]{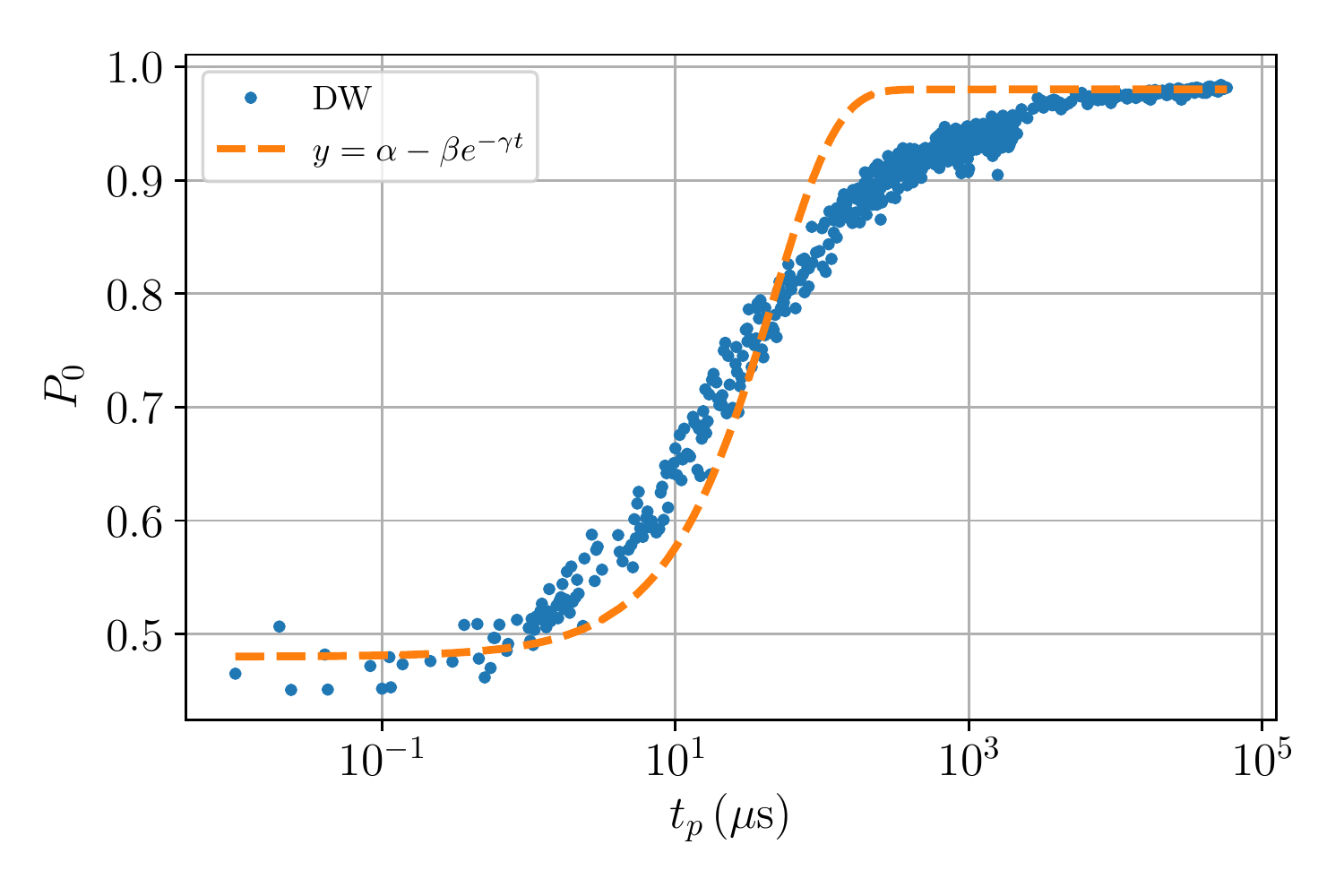}
    \includegraphics[width=0.98\columnwidth]{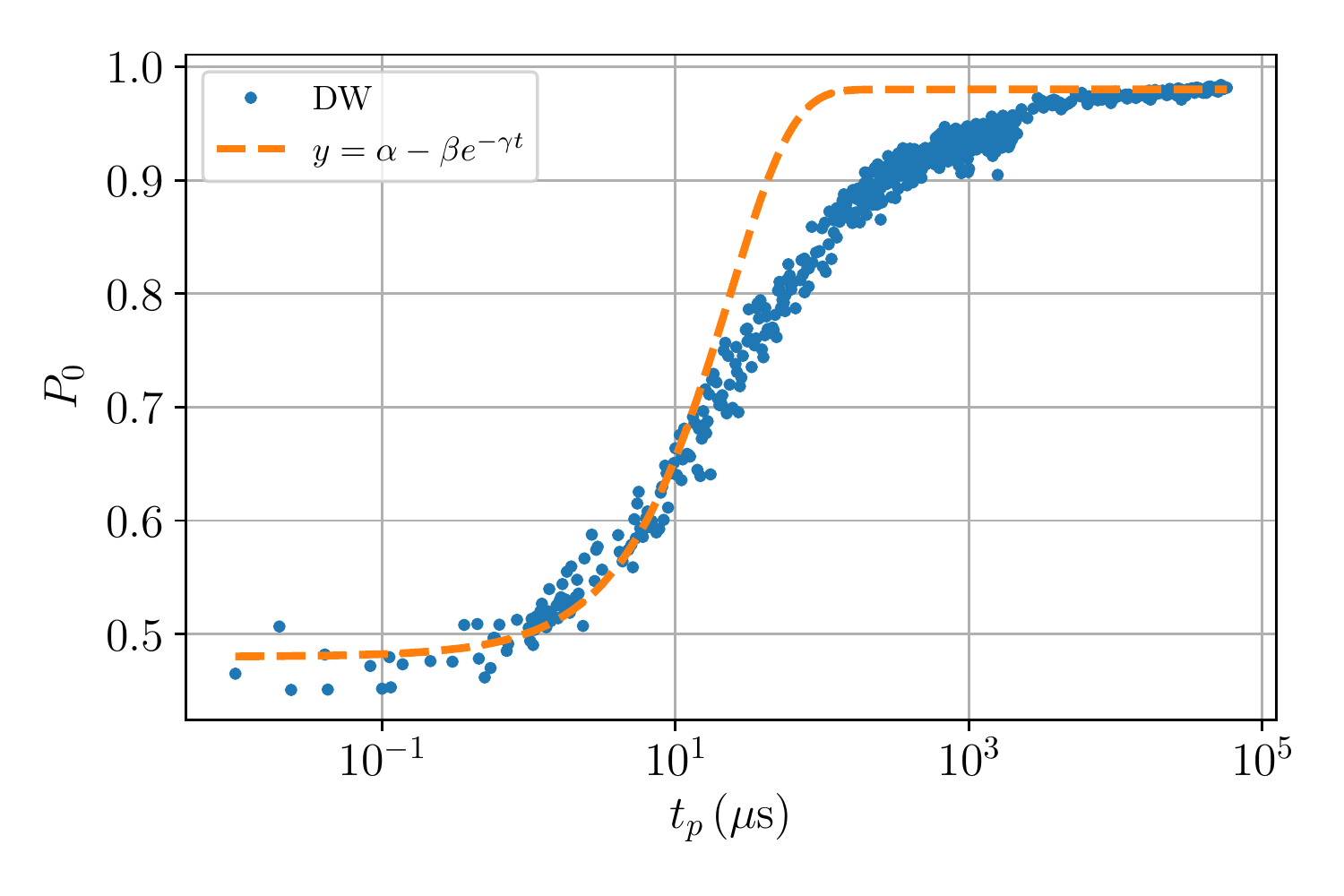}
    \caption{\textbf{Single time-scale fit for DW under the pause.} {(Top) Fixing $\alpha = P_0(t_p\rightarrow \infty)$ and $\beta =  \alpha - P_0(t_p=0)$, we obtain $\gamma^{-1} = 47\mu$s from a non-linear least squares fitting procedure. (Bottom) Using the same functional form, but replacing the single time-scale $1/\gamma$ by the dominant time-scale from Fig.~\ref{fig:P0_tp} (top).}}
    \label{fig:dw-single-time}
\end{figure}

In Fig.~\ref{fig:dw-single-time} we plot the data of Fig.~\ref{fig:P0_tp} (top) with only a single time-scale for the curve fit. {We see it does not describe the data accurately, in both cases where $\gamma$ is a free parameter, or where we fix $\gamma$ with the dominant value (0.043) found from the two time-scale fit in the main text (Fig.~\ref{fig:P0_tp} (top)). Note however, that in the latter case, for small values of $t_p \lesssim 20 \mu$s the fit matches the data reasonably well, which justifies the approximation for our single time-scale analysis in the main text (Sect.~\ref{sec:Condition}).}
Moreover, the addition of extra decay parameters does not identify other time-scales (i.e. two unique ones are identified).

\end{document}